\documentclass[twocolumn,twoside]{IEEEtran}
\ifCLASSINFOpdf
 \usepackage[pdftex]{graphicx}
\else
  \usepackage[dvips]{graphicx}
\fi

\usepackage[cmex10]{amsmath}
\usepackage{amsfonts}
\usepackage{amssymb}
\usepackage{amsthm}
\usepackage{relsize}
\usepackage{enumerate}
\usepackage{booktabs}
\usepackage{morefloats}
\usepackage{cite}
\usepackage{bm}
\usepackage{setspace}
\usepackage[caption=false,font=normalsize,labelfont=sf,textfont=sf]{subfig}
\theoremstyle{plain}
\newtheorem{theorem}{Theorem}
\newtheorem{lemma}[]{Lemma}

\newtheorem{corollary}[]{Corollary}

\begin{document}
%
\bibliographystyle{ieeetran}
\title{Maximum Sum Rate of Slotted Aloha with Capture}

\author{Yitong Li and Lin~Dai,~\IEEEmembership{Senior Member,~IEEE}

\thanks{Manuscript received April 25, 2015; revised September 28, 2015 and November 29, 2015; accepted December 5, 2015. The associate editor coordinating the review of this paper and approving it for publication was P. Popovski.}
\thanks{This work was supported by the Research Grants Council (RGC) of
Hong Kong under GRF Grant CityU 112810 and the CityU Strategic Research Grant 7004232.}
\thanks{The authors are with the Department
of Electronic Engineering, City University of Hong Kong,
83 Tat Chee Avenue, Kowloon Tong, Hong Kong, China (email: yitongli2-c@my.cityu.edu.hk, lindai@cityu.edu.hk).}}

\maketitle

\begin{abstract}
The sum rate performance of random-access networks crucially depends on the access protocol and receiver structure. Despite extensive studies, how to characterize the maximum sum rate of the simplest version of random access, Aloha, remains an open question. In this paper, a comprehensive study of the sum rate performance of slotted Aloha networks is presented. By extending the unified analytical framework proposed in \cite{Dai_Aloha, Dai_CSMA} from the classical collision model to the capture model, the network steady-state point in saturated conditions is derived as a function of the signal-to-interference-plus-noise ratio (SINR) threshold which determines a fundamental tradeoff between the information encoding rate and the network throughput. To maximize the sum rate, both the SINR threshold and backoff parameters of nodes should be properly selected. Explicit expressions of the maximum sum rate and the optimal setting are obtained, which show that similar to the sum capacity of the multiple access channel, the maximum sum rate of slotted Aloha also logarithmically increases with the mean received signal-to-noise ratio (SNR), but the high-SNR slope is only $e^{-1}$. Effects of backoff and power control on the sum rate performance of slotted Aloha networks are further discussed, which shed important light on the practical network design.

\end{abstract}



\section{Introduction}

Random access provides a simple and elegant solution for multiple users to share a common channel. Studies on random-access protocols date back to 1970s \cite{Abramson1}. After decades of extensive research, random access has found wide applications to Ethernet, IEEE 802.11 networks, Long-Term Evolution (LTE) cellular systems and wireless ad-hoc networks \cite{Kurose}. The minimum coordination and distributed control make it highly appealing for low-cost data networks.

In sharp contrast to the simplicity in concept, the performance analysis of random-access networks\footnote{Unless otherwise specified, throughout the paper we only consider synchronized slotted networks where the time is divided into multiple slots, and nodes transmit at the beginning of each time slot.} has been known as notoriously difficult, which is mainly due to the lack of a coherent analytical framework. Numerous models have been proposed based on distinct assumptions. According to the receiver structure, they can be broadly divided into three categories:

\begin{enumerate}
\item \textit{Collision model}: In the classical collision model, when multiple nodes transmit their packets simultaneously, a collision occurs and none of them can be successfully decoded. A packet transmission is successful only if there are no concurrent transmissions. The collision model was first proposed by Abramson in \cite{Abramson1}, and has been widely used since then \cite{Abramson2, Kleinrock, Carleial, Lam, Feguson, Fayolle, Tsybakov,Mikhailov,Hajek,Rivest,Rao,Anantharam,Szpankowski,Luo1,Wan,Bianchi,Qin,Dai_Aloha,Dai_CSMA}.

\item \textit{Capture model}: Though an elegant and useful simplification of the receiver, the collision model could be overly pessimistic if there exists a large difference of received power. It was first pointed out by Roberts in \cite{Roberts} that even with multiple concurrent transmissions, the strongest signal could be successfully detected as long as the signal-to-interference ratio (SIR) is high enough. It was referred to as the ``capture effect'', which has been extensively studied in \cite{Namislo,Arnbak,Goodman,Habbab,Sheikh,Plas,Zorzi,Peh,Angel,Luo2,Rasool,Wieselthier,Naware1,Dua}. With the capture model, each node's packet is decoded independently by treating others' as background noise. A packet can be successfully decoded as long as its received signal-to-interference-plus-noise ratio (SINR) is above a certain threshold. It is clear that multiple packets may be successfully decoded if the SINR threshold is sufficiently low.

\item \textit{Joint-decoding model}:  Both the collision model and the capture model are essentially single-user detectors. Multiuser detectors, such as Minimum Mean Square Error (MMSE) and Successive Interference Cancelation (SIC), have been also applied to random-access networks \cite{Shinomiya,Medard,Adireddy,Naware2,Yim,Minero,Zanella}. By jointly decoding multiple nodes' packets, the efficiency can be greatly improved, though at the cost of increased receiver complexity.
\end{enumerate}


Note that the capture model and the joint-decoding model both have the so-called ``multipacket-reception (MPR)'' capability \cite{Ghez1,Ghez2}, and have been referred to as the MPR model in many references \cite{Luo2,Dua,Shinomiya,Adireddy,Naware2,Yim,Zanella}. 
Here we distinguish them apart because they assume different receiver structures.
In this paper, we specifically focus on the performance analysis based on the capture model.

\subsection{Maximum Network Throughput of Slotted Aloha}

In random-access networks, due to the uncoordinated nature of transmitters, the number of successfully decoded packets in each time slot varies from time to time. In the literature, the average number of successfully decoded packets per time slot is usually adopted as an important performance metric, which is referred to as the network throughput.

The network throughput performance depends on a series of key factors including the receiver model and protocol design. With the classical collision model, for instance, at most one packet can be successfully decoded at each time slot. Therefore, the network throughput, which is also the fraction of time that an effective output is produced in this case, cannot exceed $1$. The maximum network throughput of slotted Aloha was shown to be only $e^{-1}$ with the collision model \cite{Abramson2}, which indicates that over $60\%$ of the time is wasted when the network is either in collision or idle states. To improve the efficiency, Carrier Sense Multiple Access (CSMA) was further introduced in \cite{Kleinrock}, with which the network throughput can approach $1$ by reducing the sensing time. On the other hand, significant improvement in network throughput was also observed when the capture model is adopted \cite{Namislo,Arnbak,Goodman,Habbab,Sheikh,Zorzi,Peh,Luo2,Rasool,Dua}.
Intuitively, with the capture model, more packets can be successfully decoded by reducing the SINR threshold. The network throughput is thus greatly improved, and may exceed $1$ if the SINR threshold is sufficiently small.

Despite extensive studies, how to maximize the network throughput has been an open question for a long time. In Abramson's landmark paper \cite{Abramson2}, by modeling the aggregate traffic as a Poisson random variable with parameter $G$, the network throughput of slotted Aloha with the collision model can be easily obtained as $Ge^{-G}$, which is maximized at $e^{-1}$ when $G=1$. To enable the network to operate at the optimum point for maximum network throughput, nevertheless, it requires the connection between the mean traffic rate $G$ and key system parameters such as transmission probabilities of nodes, which turns out to be a challenging issue.
Various retransmission strategies were developed to adjust the transmission probability of each node according to the number of backlogged nodes to stabilize\footnote{Note that various definitions of stability have been developed in the literature. A widely adopted one is that a network is stable if the network throughput is equal to the aggregate input rate.} the network \cite{Carleial,Lam, Feguson, Fayolle}. Yet most of them were based on the realtime feedback information on the backlog size, which may not be available in a distributed network. Decentralized retransmission control was further studied in \cite{Lam,Mikhailov,Hajek,Rivest}, where algorithms were proposed to either estimate and feed back the backlog size \cite{Mikhailov,Rivest}, or update the transmission probability of each node recursively according to the channel output \cite{Lam, Hajek}.

The above analytical approaches were also applied to the capture model. By assuming Poisson distributed aggregate traffic, for instance, the network throughput was derived as a function of the mean traffic rate $G$ and the SIR threshold in \cite{Arnbak,Goodman,Sheikh} under distinct assumptions on channel conditions. Similar to the case of collision model, the maximum network throughput can be obtained by optimizing $G$, yet how to properly tune the system parameters to achieve the maximum network throughput remains unknown. Retransmission control strategies developed in \cite{Carleial}, \cite{Mikhailov} and \cite{Hajek} were further extended to the capture model in \cite{Plas}, \cite{Peh} and \cite{Zorzi}, respectively. To evaluate the network throughput performance for given transmission probabilities of nodes, various Markov chains were also established in \cite{Namislo,Habbab,Rasool,Dua} to model the state transition of each individual user. The computational complexity, nevertheless, sharply increases when sophisticated backoff strategies are further involved, which renders it extremely difficult, if not impossible, to search for the optimal configuration to maximize the network throughput.

The difficulty originates from the modeling of random-access networks. As demonstrated in \cite{Dai_CSMA}, the modeling approaches in the literature can be roughly divided into two categories: channel-centric \cite{Abramson2, Kleinrock, Carleial, Lam, Feguson, Fayolle,Mikhailov,Hajek,Rivest} and node-centric \cite{Tsybakov,Rao,Anantharam,Szpankowski,Luo1,Wan,Bianchi}. By focusing on the state transition process of the aggregate traffic, the channel-centric approaches capture the essence of contention among nodes, which, nevertheless, ignore the behavior of each node's queue and thus shed little light on the effect of backoff parameters on the performance of each single node. With the node-centric approaches, on the other hand, the modeling complexity becomes prohibitively high if interactions among nodes' queues are further taken into consideration. To simplify the analysis, a key approximation, which has been widely adopted and shown to be accurate for performance evaluation of large multi-queue systems \cite{Hui}, is to treat each node's queue as an independent queueing system with identically distributed service time. The service time distribution is still crucially determined by the aggregate activities of head-of-line (HOL) packets of all the nodes, which requires proper modeling of HOL packets' behavior.

In our recent work \cite{Dai_Aloha,Dai_CSMA}, a unified analytical framework for two representative random-access protocols, Aloha and CSMA, was established, where the network steady-state points were characterized based on the fixed-point equations of the limiting probability of successful transmission of HOL packets by assuming the classical collision model. As we will show in this paper, the proposed analytical framework can be further extended to incorporate the capture model, based on which explicit expressions of the maximum network throughput and the corresponding optimal transmission probabilities of nodes will be derived.

\subsection{Maximum Sum Rate of Slotted Aloha}

From the information-theoretic perspective, random access can be regarded as a multiple access channel (MAC) with a random number of active transmitters. It is well known that the sum capacity of an $n$-user Additive-White-Gaussian-Noise (AWGN) MAC is determined by the received SNRs, i.e., $C_{sum}=\log_{2}(1+\sum_{i=1}^{n}SNR_i)$. With random access, however, the number of active transmitters is a random variable whose distribution is determined by the protocol and parameter setting. Moreover, to achieve the sum capacity, a joint decoding of all transmitted codewords should be performed at the receiver side, which might be unaffordable for random-access networks. Therefore, the sum rate performance of random access becomes closely dependent on assumptions on the access protocol and receiver design.
\begin{table*}[!tp]\small\label{Table:main_notation}
  \begin{center}
    \caption{Main Notations}
    \begin{tabular}{cl}
      \toprule
      $n$ & Number of nodes \\
      $\rho$ & Mean received SNR \\
      $\mu$ & SINR threshold\\
      $R$ & Information encoding rate of nodes\\
      $\hat{\lambda}_{out}$ & Network throughput\\
      $p$ & Steady-state probability of successful transmission of HOL packets\\
      $K$ & Cutoff phase of HOL packets \\
      $\{q_i\}_{i=0, \ldots, K}$ & Transmission probabilities of nodes \\
      $\hat{\lambda}_{\max}$ & Maximum network throughput\\
      $C$ & Maximum sum rate\\
      \bottomrule
    \end{tabular}
  \end{center}
\end{table*}

There has been a great deal of effort to explore the information-theoretic limit of random-access networks. For instance, the concept of rate splitting \cite{Rimoldi} was first introduced to slotted Aloha networks in \cite{Medard}, where a joint coding scheme was developed for the two-node case. If each node independently encodes its information, \cite{Minero} showed that the sum rate\footnote{Note that different terminologies were used in these studies. In \cite{Angel}, for instance, ``average spectral efficiency'' was used to denote the sum rate of slotted Aloha. In \cite{Minero,Naware1,Qin, Wieselthier}, it was referred to as ``throughput''.} performance of slotted Aloha networks can be improved by adaptively adjusting the encoding rate according to the number of nodes and the transmission probability of each node. \cite{Medard} and \cite{Minero} are based on the assumption of joint decoding of multiple nodes' packets at the receiver side. With the capture model, the effects of power allocation and modulation on the sum rate of slotted Aloha in AWGN channels were analyzed in \cite{Angel} and \cite{Wieselthier}, respectively. Queueing stability and channel fading were further considered in \cite{Naware1}, where the sum rates with various cross-layer approaches were derived. In \cite{Qin}, by assuming that each node has its own channel state information (CSI) and the collision model is adopted at the receiver side, the scaling behavior of the sum rate of slotted Aloha as the number of nodes $n$ goes to infinity was characterized, and shown to be identical to that of the sum capacity of MAC.



Although various analytical models were developed in the above studies, many of them rely on numerical methods to calculate the sum rate under specific settings. It remains largely unknown how to maximize the sum rate by optimizing the system parameters. As we will demonstrate in this paper, the sum rate optimization of slotted Aloha networks can be decomposed into two parts: 1) For given information encoding rate $R$, or equivalently, SINR threshold $\mu$, the network throughput can be maximized by properly choosing backoff parameters, i.e., the transmission probabilities of nodes. 2) As the information encoding rate and the maximum network throughput are both functions of the SINR threshold $\mu$, the sum rate can be further optimized by tuning $\mu$.

Specifically, we characterize the maximum sum rate of slotted Aloha with the capture model by considering an $n$-node slotted Aloha network where all the nodes transmit to a single receiver with the SINR threshold $\mu$, and the received SNRs of nodes' packets are assumed to be exponentially distributed with the mean received SNR $\rho$. The main findings are summarized below.
\begin{enumerate}
\item The network steady-state point in saturated conditions, which is characterized as the single non-zero root of the fixed-point equation of the limiting probability of successful transmission of HOL packets, is found to be closely dependent on transmission probabilities of nodes, the SINR threshold $\mu$ and the mean received SNR $\rho$.
\item The maximum sum rate is derived as a function of the mean received SNR $\rho$. Similar to the sum capacity of MAC, it also logarithmically increases with $\rho$, but the high-SNR slope is only $e^{-1}$. In the low SNR region, it is a monotonic increasing function of the number of nodes $n$, and approaches $e^{-1}\log_2 e\approx 0.5307$ as $n\to\infty$.
\item To achieve the maximum sum rate, both the SINR threshold and the transmission probabilities of nodes should be carefully selected according to the mean received SNR $\rho$. Explicit expressions of the optimal SINR threshold and transmission probabilities are derived, and verified by simulations.
\end{enumerate}

Note that the MAC scenario considered in this paper should be distinguished from the ad-hoc scenario which has been extensively studied in recent years \cite{Bacelli_2006, Ganti_2009, Weber_2010, Nardelli_2012, Bacelli_2013, Zhong_2014, George_2015}. In contrast to the MAC where multiple nodes transmit to a common receiver, multiple transmitter-receiver pairs exist in the ad-hoc case. Representative applications of the former one include cellular systems and IEEE 802.11 networks, where in each cell/basic-service-set, multiple users transmit to the base-station/access-point. The latter is usually considered in a wireless ad-hoc network, such as wireless sensor networks.

The remainder of this paper is organized as follows. Section II presents the system model. Section III focuses on the network throughput analysis, where the maximum network throughput and the optimal backoff parameters are obtained as functions of the SINR threshold and the mean received SNR. The maximum sum rate is derived in Section IV, and simulation results are presented in Section V. The effects of key factors, including backoff and power control, are discussed in Section VI. Conclusions are summarized in Section VII. Table I lists the main notations used in this paper.

\section{System Model}\label{Section2}

\begin{figure}[htp]
\centering
\includegraphics[width=3.5in,height=1.8in]{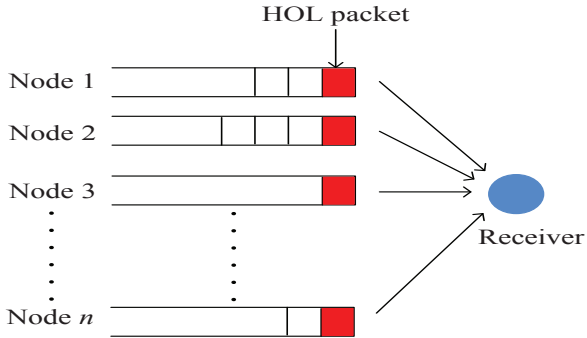}
\caption{Graphic illustration of an $n$-node slotted Aloha network.}
\label{MQSS}
\end{figure}

Consider a slotted Aloha network where $n$ nodes transmit to a single receiver, as Fig. \ref{MQSS} illustrates.  All the nodes are synchronized and can start a transmission only at the beginning of a time slot. For each node, assume that it always has packets in its buffer and each packet transmission lasts for one time slot. We assume perfect and instant feedback from the receiver and ignore the subtleties of the physical layer such as the switching time from receiving mode to transmitting mode and the delay required for information exchange.

\begin{figure}[htb]
\centering
\includegraphics[width=3.5in,height=1.5in]{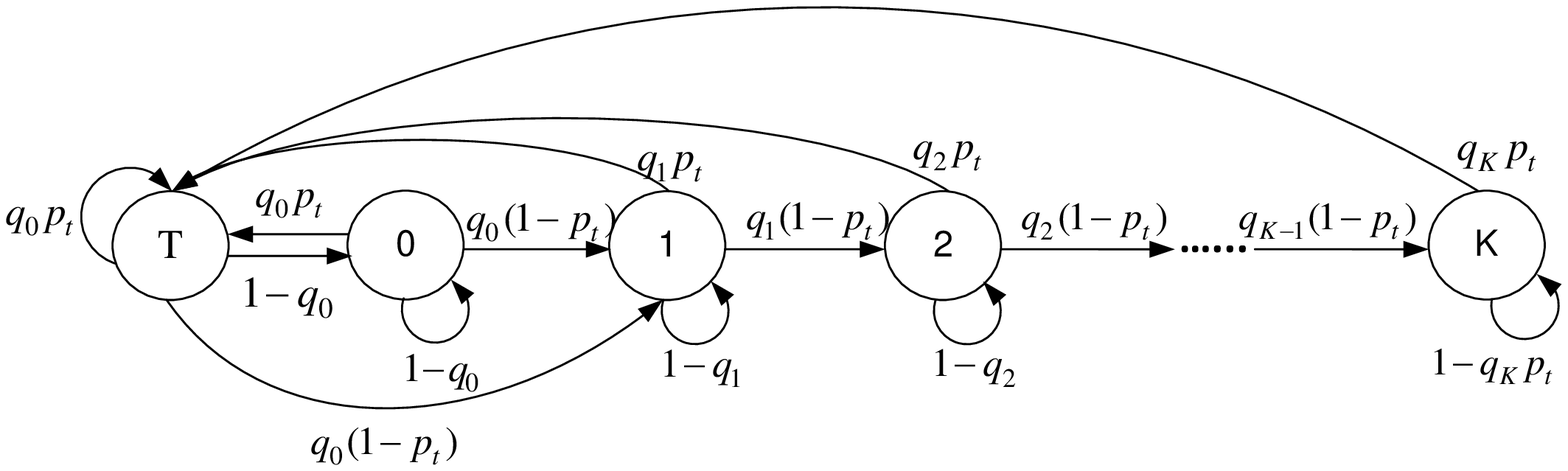}
\caption{State transition diagram of an individual HOL packet in slotted Aloha networks.}
\label{markov}
\end{figure}

Let $g_k$ denote the channel gain from node $k$ to the receiver, which can be further written as $g_k=\gamma_k\cdot h_k$.
$h_k$ is the small-scale fading coefficient of node $k$ which varies from time slot to time slot\footnote{More specifically, we assume that the time slot length is equal to the channel coherence time.} and is modeled as a complex Gaussian random variable with zero mean and unit variance. The large-scale fading coefficient $\gamma_k$ characterizes the long-term channel effect such as path loss and shadowing. Due to the slow-varying nature, the large-scale fading coefficients are usually available at the transmitter side through channel measurement. Let us first assume that power control is performed to overcome the effect of large-scale fading.\footnote{In practical systems such as cellular systems, the base-station sends a pilot signal periodically for all the users in its cell to measure their large-scale fading gains and adjust their transmission power accordingly to maintain constant mean received power. This process is usually referred to as open-loop power control. It should also be noted that in the ad-hoc scenario, the difference in the large-scale fading gains from a certain node and its interferers cannot be removed by power control as they may transmit to different receivers. In that case, nodes would have distinct mean received SNRs which are closely dependent on their spacial locations.} Specifically, denote the transmission power of node $k$ as $\bar{P}_k$. Then we have
\begin{equation}
\label{power_control}
\bar{P}_k \cdot |\gamma_k|^2=P_0.
\end{equation}
In this case, each node has the same mean received SNR
$\rho=P_0/\sigma^2$. The assumption of power control will be relaxed in Section \ref{Section5-3}, where the analysis is extended to incorporate distinct mean received SNRs.

Throughout the paper, we assume that the receiver always has perfect channel state information but the transmitters are unaware of the instantaneous realizations of the small-scale fading coefficients. As a result, each node independently encodes its information at a given rate $R$ bit/s/Hz. Assume that each codeword lasts for one time slot,\footnote{Note that here we assume that each codeword only covers one channel coherence time period. Without coding over fading states, the decoding delay is greatly reduced, but a certain rate loss is caused, as we will show in Section \ref{Section4-2} and Section \ref{Section5-1}. Recent studies have also shown that significant gains can be achieved by introducing coding over successive packets of each node \cite{Casini_2007, Liva_2011, Stefanovic_2013, Goseling_2015}, which is referred to as ``coded random access" \cite{Paolini_2015}.} and the capture model is adopted at the receiver side. That is, no joint decoding is performed among nodes' packets or with previously received packets. Instead, each node's packet is decoded independently by treating others' as background noise at each time slot, and a packet can be successfully decoded if its received signal-to-interference-plus-noise ratio (SINR) is above a certain threshold.

Let
\begin{equation}
\label{mu}
\mu=2^R-1
\end{equation}
denote the SINR threshold at the receiver. For each node's packet, if its received SINR exceeds the threshold $\mu$, it can be successfully decoded and rate $R$ can be supported for reliable communications.\footnote{More specifically, denote the received SINR of node $k$ as ${\eta_k}$. If $\log_2(1+{\eta_k})>R$, then by random coding the error probability of node $k$'s packet is exponentially reduced to zero as the block length goes to infinity. Here we assume that the block length is sufficiently large such that node $k$'s packet can be successfully decoded as long as ${\eta_k}\geq \mu$. Note that this is an ideal case. In practice, the threshold not only depends on the information encoding rate $R$, but also the error probability that is determined by the coding and decoding schemes.} Note that when the SINR threshold $\mu$ is sufficiently small, more than one packets could be successfully decoded at each time slot. It is clear that the number of successfully decoded packets in time slot $t$, denoted by $N_t$, is a time-varying variable. As a result, the total received information rate, i.e., $R\cdot N_t$ bit/s/Hz, also varies with time. In this paper, we focus on the long-term system behavior and define the sum rate as the time average of the received information rate:
\begin{align}\label{sum rate}
R_s=\lim_{t\to\infty} \frac{1}{t}\sum_{i=1}^{t} R\cdot N_i=R\cdot \hat{\lambda}_{out},
\end{align}
where
\begin{align}\label{network_throughput_definition}
\hat{\lambda}_{out}=\lim_{t\to\infty} \frac{1}{t}\sum_{i=1}^{t}N_i
\end{align}
is the average number of successfully decoded packets per time slot, which is referred to as the network throughput.

Both the information encoding rate $R$ and the network throughput $\hat{\lambda}_{out}$ depend on the SINR threshold $\mu$. Intuitively, by reducing $\mu$, more packets can be successfully decoded at each time slot, yet the information encoding rate becomes smaller. Therefore, the SINR threshold $\mu$ should be carefully chosen to maximize the sum rate. Note that the network throughput $\hat{\lambda}_{out}$ is also crucially determined by the protocol design and backoff parameters. In the next section, we will specifically focus on the network throughput performance of slotted Aloha networks.

\section{Network Throughput}\label{Section3}

As Fig. \ref{MQSS} illustrates, an $n$-node buffered slotted Aloha network is essentially an $n$-queue-single-server system whose performance is determined by the aggregate activities of HOL packets. In this section, we will first characterize the state transition process of HOL packets, and then derive the network steady-state point in saturated conditions as the single non-zero root of the fixed-point equation of the steady-state probability of successful transmission of HOL packets. Finally, the maximum network throughput will be obtained by optimizing the transmission probabilities of nodes.

\subsection{State Characterization of HOL Packets}\label{Section3-1}
The behavior of each HOL packet can be modeled as a discrete-time Markov process. As Fig. \ref{markov} shows,\footnote{Note that a similar Markov chain of the HOL packet was established in \cite{Dai_Aloha} where the transmission probability of each fresh HOL packet was assumed to be $1$. Here the original State 0 is split into two states, i.e., State T and State 0, to incorporate a general transmission probability of $0<q_0\leq 1$ for each fresh HOL packet.} a fresh HOL packet is initially in State T, and moves to State 0 if it is not transmitted. Define the phase of a HOL packet as the number of collisions it experiences. A phase-$i$ HOL packet stays in State $i$ if it is not transmitted. Otherwise, it moves to State T if its transmission is successful, or State $\min(K,i+1)$ if the transmission fails, where $K$ denotes the cutoff phase. Note that the cutoff phase $K$ can be any non-negative integer. When $K=0$, States 0 and $K$ in Fig. \ref{markov} would be merged into one state, i.e., State 0.
Intuitively, to alleviate the contention, nodes should reduce their transmission probabilities as they experience more collisions. Therefore, we assume that the transmission probabilities $\{ q_{i} \} _{i=0,...,K} $ form a monotonic non-increasing sequence.

In Fig. \ref{markov}, $p_t$ denotes the probability of successful transmission of HOL packets at time slot $t$.\footnote{Note that in Fig. \ref{markov}, the probability of successful transmission of HOL packets at time slot $t$, $p_t$, is assumed to be state independent. Intuitively, given that a HOL packet is attempting to transmit, the probability that its transmission is successful is determined by the overall activities of all the other HOL packets, rather than its own state. Therefore, no matter which state the HOL packet is currently staying at, its probability of successful transmission only depends on the attempt rate of other HOL packets at the moment, which is denoted as $p_t$ in Fig. \ref{markov}.} It can be easily shown that the Markov chain is uniformly strongly ergodic if and only if the limit $\mathop{\lim }\limits_{t\to \infty } p_{t} =p$
exists \cite{Iosifescu}. The steady-state probability distribution $\{\pi_i\}$ of the Markov chain in Fig. \ref{markov} can be further obtained as
\begin{equation}
\pi_T=\tfrac{1}{\sum_{i=0}^{K-1}\tfrac{\left(1-p\right)^i}{q_i}+\tfrac{\left(1-p\right)^K}{pq_K}},
\label{pi0}
\end{equation}
and
\begin{equation}
\begin{cases}
\pi_0=\tfrac{1-pq_0}{pq_0}\pi_T.  & K=0 \\
\pi_0=\tfrac{1-q_0}{q_0}\pi_T, \;\pi_i=\tfrac{(1-p)^i}{q_i}\pi_T, \;i=1,\ldots,K-1, \\
\pi_K=\tfrac{(1-p)^K}{pq_K}\pi_T.  & K\geq 1
\end{cases}
\label{piK}
\end{equation}

Note that $\pi_T$ is the service rate of each node's queue as the queue has a successful output if and only if the HOL packet is in State T.

\subsection{Steady-state Point in Saturated Conditions}\label{Section3-2}
By regarding an $n$-node buffered slotted Aloha network as an $n$-queue-single-server system, we can see that the network throughput $\hat{\lambda}_{out}$ is indeed the system output rate, which is equal to the aggregate input rate $\hat{\lambda}$ if each node's buffer has a non-zero probability of being empty. As $\hat{\lambda}$ increases, the network will eventually become saturated where each node is busy with a non-empty queue. In this case, the network throughput is determined by the aggregate service rate, i.e.,
\begin{equation} \label{throughput}
\hat{\lambda}_{out}=n\pi_T,
\end{equation}
which, as (\ref{pi0}) shows, depends on the steady-state probability of successful transmission of HOL packets $p$. In this section, we will characterize the network steady-state point in saturated conditions based on the fixed-point equation of $p$.

Specifically, for HOL packet $j$, let $\mathcal{S}_j$ denote the set of nodes which have concurrent transmissions. It can be successfully decoded at the receiver side if and only if its received SINR is above the threshold $\mu$, i.e., $\tfrac{P_j}{\sum_{k\in \mathcal{S}_j}P_k+\sigma^2}\geq \mu$, where $P_k=\bar{P}_k|g_k|^2=P_0|h_k|^2$ denotes the received power according to (\ref{power_control}). Suppose that $|\mathcal{S}_j|=i$. The steady-state probability of successful transmission of HOL packet $j$ given that there are $i$ concurrent transmissions, $r_i^j$, can be then written as
\begin{equation}\label{rij}
r_i^j=\textrm{Pr}\left\{\tfrac{|h_j|^2}{\sum_{k \in \mathcal{S}_j  } |h_k|^2+\tfrac{1}{\rho}}\geq\mu\right\},
\end{equation}
where $\rho=P_0/\sigma^2$ is the mean received SNR. With $h_k\sim \mathcal{CN}(0,1)$, $r_i^j$ can be easily obtained as \cite{Arnbak,Dua}
\begin{equation}\label{ri1}
r_i^j=\tfrac{\exp\left(-\tfrac{\mu}{\rho}\right)}{(\mu+1)^i}.
\end{equation}
The right-hand side of (\ref{ri1}) is independent of $j$, indicating that all the HOL packets have the same conditional probability of successful transmission.\footnote{Note that here all the HOL packets have the same conditional probability of successful transmission because their mean received SNRs are assumed to be equal.} Therefore, we drop the superscript $j$, and write the steady-state probability of successful transmission of HOL packets $p$ as
\begin{equation}\label{Probability-of-success_1}
p=\sum_{i=0}^{n-1} r_i \cdot \text{Pr\{$i$ concurrent transmissions\}}.
\end{equation}

In saturated conditions, all the nodes have non-empty queues. According to the Markov chain shown in Fig. \ref{markov}, the probability that the HOL packet is requesting transmission is given by $\pi_T q_0+\sum_{i=0}^K \pi_i q_i$, which is equal to $\pi_T/p$ according to (\ref{piK}). Therefore, the probability that there are $i$ concurrent transmissions can be obtained as
\begin{align}\label{Probability-of-success_2}
&\text{Pr\{$i$ concurrent transmissions\}}   \notag \\
&=\binom{n-1}{i}\left(1-\tfrac{\pi_T}{p}\right)^{n-1-i} \left(\tfrac{\pi_T}{p}\right)^{i}.
\end{align}
By substituting \eqref{ri1} and \eqref{Probability-of-success_2} into \eqref{Probability-of-success_1}, the steady-state probability of successful transmission of HOL packets $p$ can be obtained as
\begin{align}\label{Probability-of-success_3}
p&=\exp\left(-\tfrac{\mu}{\rho}\right)\cdot\left(1-\tfrac{\mu}{\mu+1}\cdot \tfrac{\pi_T}{p}\right)^{n-1} \notag \\
&\stackrel{\text{for large } n}{\approx} \exp\left\{-\tfrac{\mu}{\rho}-\tfrac{n\mu}{\mu+1}\cdot \tfrac{\pi_T}{p} \right\},
\end{align}
\begin{figure*}[!tp]
\centering
\subfloat[]{
\label{maxThroughput-threshold}
\includegraphics[width=3.5in,height=2.2in]{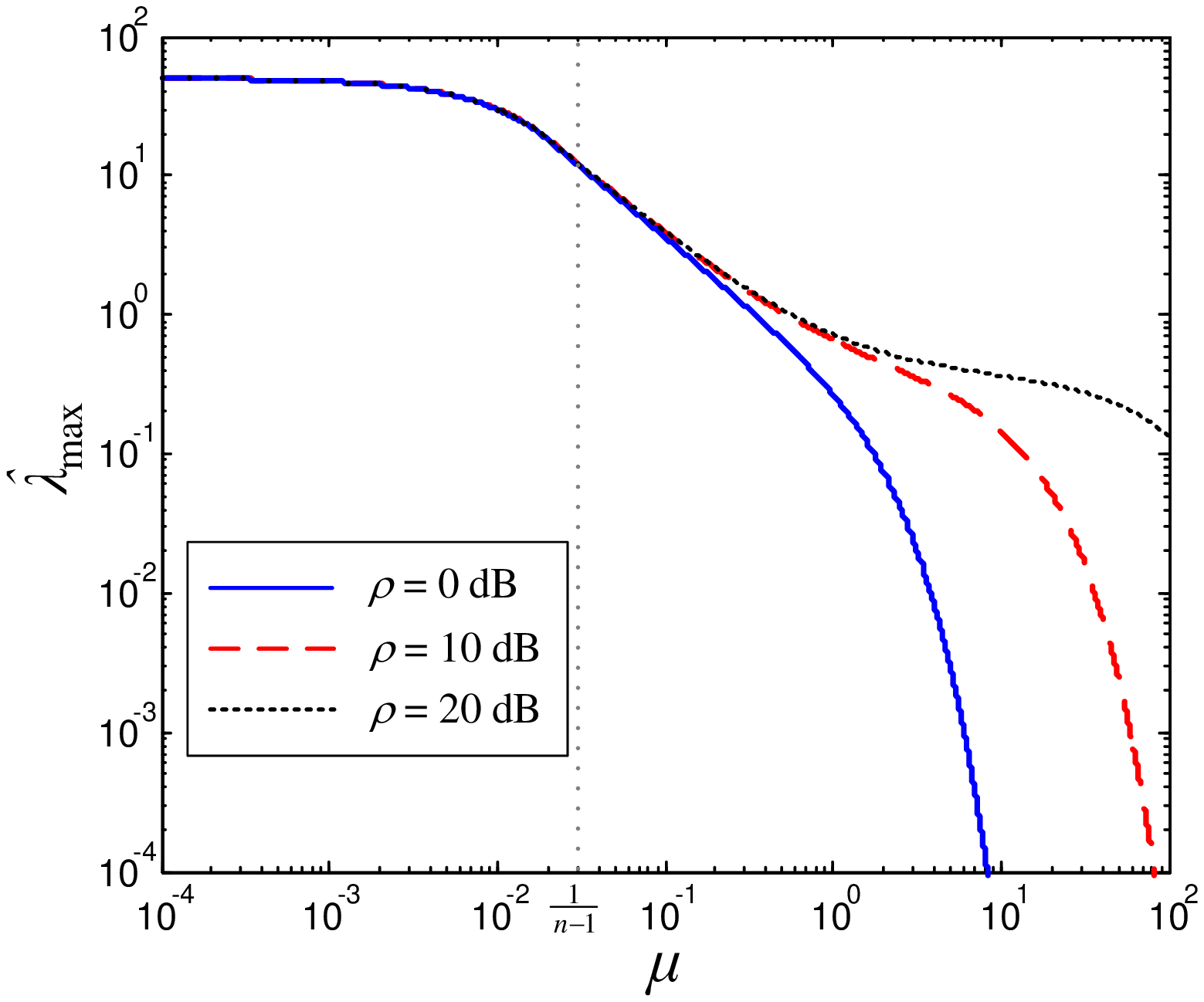}}
\hfil
\subfloat[]{
\label{maxThroughput-SNR}
\includegraphics[width=3.5in,height=2.2in]{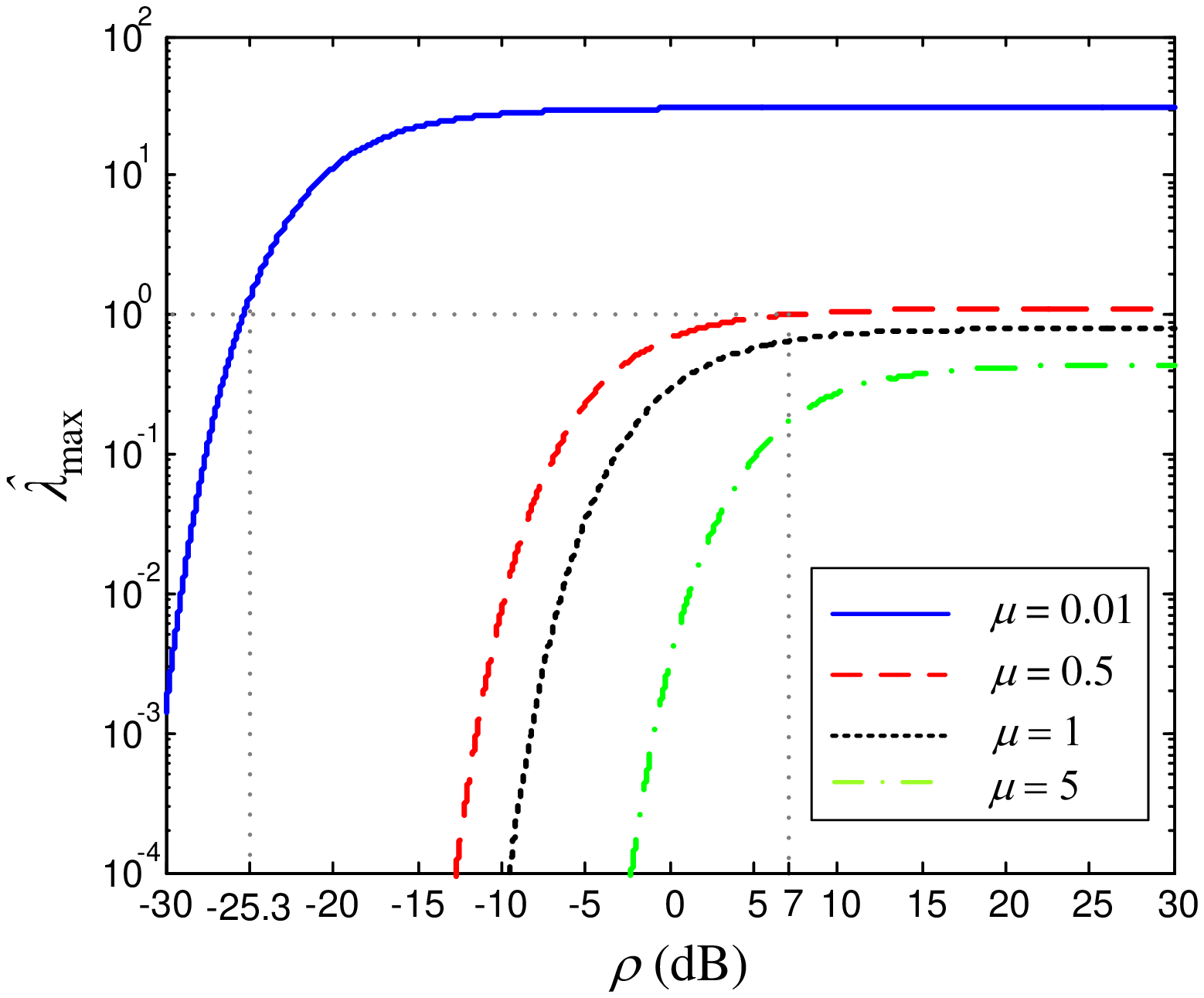}}
\caption{Maximum network throughput $\hat{\lambda}_{\max}$ versus (a) SINR threshold $\mu$ and (b) mean received SNR $\rho$. $n=50$.}
\label{Fig_maxThroughput}
\end{figure*}
where the approximation is obtained by applying $(1-x)^n\approx \exp{(-nx)}$ for $0<x<1$.\footnote{Note that with a small network size, i.e., $n\le 5$ for instance, the approximation error may become noticeable. It, nevertheless, rapidly declines as the number of nodes $n$ increases.} Finally, by substituting \eqref{pi0} into \eqref{Probability-of-success_3}, we have
\begin{equation}\label{Probability-of-success_4}
p\approx \exp\left(-\tfrac{\mu}{\rho}-\tfrac{n\mu}{\mu+1}\cdot \tfrac{1}{\resizebox{5mm}{!}{$\sum$}_{i=0}^{K-1} \tfrac{p\left(1-p\right)^i}{q_i}+\tfrac{\left(1-p\right)^K}{q_K}}\right) .
\end{equation}

The following theorem states the existence and uniqueness of the root of the fixed-point equation \eqref{Probability-of-success_4}.
\begin{theorem} \label{Theorem_pA}
The fixed-point equation \eqref{Probability-of-success_4} has one single non-zero root $p_A$ if $\{q_i\}_{i=0, \ldots, K}$ is a monotonic non-increasing sequence.
\end{theorem}
\begin{proof}
See Appendix \ref{Proof_Theorem_pA}.
\end{proof}

\noindent As we can see from (\ref{Probability-of-success_4}), the non-zero root $p_A$ is closely dependent on backoff parameters $\{q_i\}_{i=0, \ldots, K}$. Without loss of generality, let $q_i=q_0\cdot \mathcal{Q}_i$ where $q_0$ is the initial transmission probability and $\mathcal{Q}_i$ is an arbitrary monotonic non-increasing function of $i$ with $\mathcal{Q}_0=1$ and $\mathcal{Q}_{i}\leq \mathcal{Q}_{i-1}$, $i=1,\ldots,K$.
With the cutoff phase $K=0$, or the backoff function $\mathcal{Q}_i=1$, $i=0,\ldots,K$, for instance, $p_A$ can be explicitly written as
\begin{equation}\label{pA_0}
p_A=\exp\left(-\tfrac{\mu}{\rho}-\tfrac{n\mu}{\mu+1}q_0\right).
\end{equation}

\subsection{Maximum Network Throughput for Given $\mu$ and $\rho$}\label{Section3-3}
It has been shown in Section \ref{Section3-2} that the network operates at the steady-state point $p_A$ in saturated conditions. By combining (\ref{throughput}) and (\ref{Probability-of-success_3}), the network throughput at $p_A$ can be written as
\begin{equation}\label{eq15}
\hat{\lambda}_{out}=(\mu+1)\cdot\left(\tfrac{-p_A \ln p_A}{\mu}-\tfrac{p_A}{\rho}\right),
\end{equation}
where $p_A$ is an implicit function of the transmission probabilities $q_i$, $i=0, \ldots,K$, which is given in (\ref{Probability-of-success_4}). It can be seen from (\ref{eq15}) and (\ref{Probability-of-success_4}) that the network throughput is crucially determined by the backoff parameters $\{q_i\}$. In this section, we focus on the maximum network throughput $\hat{\lambda}_{\max}={\max}_{\{q_i\}}\hat{\lambda}_{out}$. The following theorem presents the maximum network throughput $\hat{\lambda}_{\max}$ and the corresponding optimal backoff parameters $\{q_i^{*}\}$.

\begin{theorem}\label{Theorem_maxThroughput}
For given SINR threshold $\mu\in(0,\infty)$ and mean received SNR $\rho\in(0,\infty)$, the maximum network throughput is given by
\begin{equation}\label{maxThroughput_all}
\hat{\lambda}_{\max}=\begin{cases}
\tfrac{\mu+1}{\mu}\exp\left(-1-\tfrac{\mu}{\rho}\right) & \text{if\;\;}\mu\geq\tfrac{1}{n-1} \\
n\exp\left(-\tfrac{n\mu}{\mu+1}-\tfrac{\mu}{\rho}\right)  & \text{otherwise},
\end{cases}
\end{equation}
which is achieved at
\begin{equation}\label{maxThroughput_q}
q_i^*=\begin{cases}
\hat{q}_0\mathcal{Q}_i & \text{if\;\;}\mu\geq\tfrac{1}{n-1} \\
1  & \text{otherwise},
\end{cases}
\end{equation}
$i=0,\ldots,K$, where $\hat{q}_0$ is given by
\begin{align}\label{qm}
\hat{q}_0&=\tfrac{\mu+1}{n\mu}\cdot \left\{ \sum_{i=0}^{K-1}\tfrac{\exp\left(-1-\tfrac{\mu}{\rho}\right)\left[1-\exp\left(-1-\tfrac{\mu}{\rho}\right)\right]^i}{\mathcal{Q}_i} \right.\notag \\
&\left.+\tfrac{\left[1-\exp\left(-1-\tfrac{\mu}{\rho}\right)\right]^K}{\mathcal{Q}_K}\right\}.
\end{align}
\end{theorem}
\begin{proof}
See Appendix \ref{Proof_Theorem_maxThroughput}.
\end{proof}

Eq. (\ref{maxThroughput_all}) shows that for given SINR threshold $\mu$, the maximum network throughput $\hat{\lambda}_{\max}$ is a monotonic increasing function of the mean received SNR $\rho$. As $\rho\to\infty$, we have
\begin{equation}
\lim_{\rho \rightarrow \infty}\hat{\lambda}_{\max}=\begin{cases}
\tfrac{\mu+1}{\mu}e^{-1} & \text{if\;\;}\mu\geq\tfrac{1}{n-1} \\
n\exp\left(-\tfrac{n\mu}{\mu+1}\right)  & \text{otherwise},
\end{cases}
\end{equation}
which approaches $e^{-1}$ when $\mu\gg1$.

\begin{figure*}[!tp]
\centering
\subfloat[]{
\label{optimalThreshold}
\includegraphics[width=3.5in,height=2.2in]{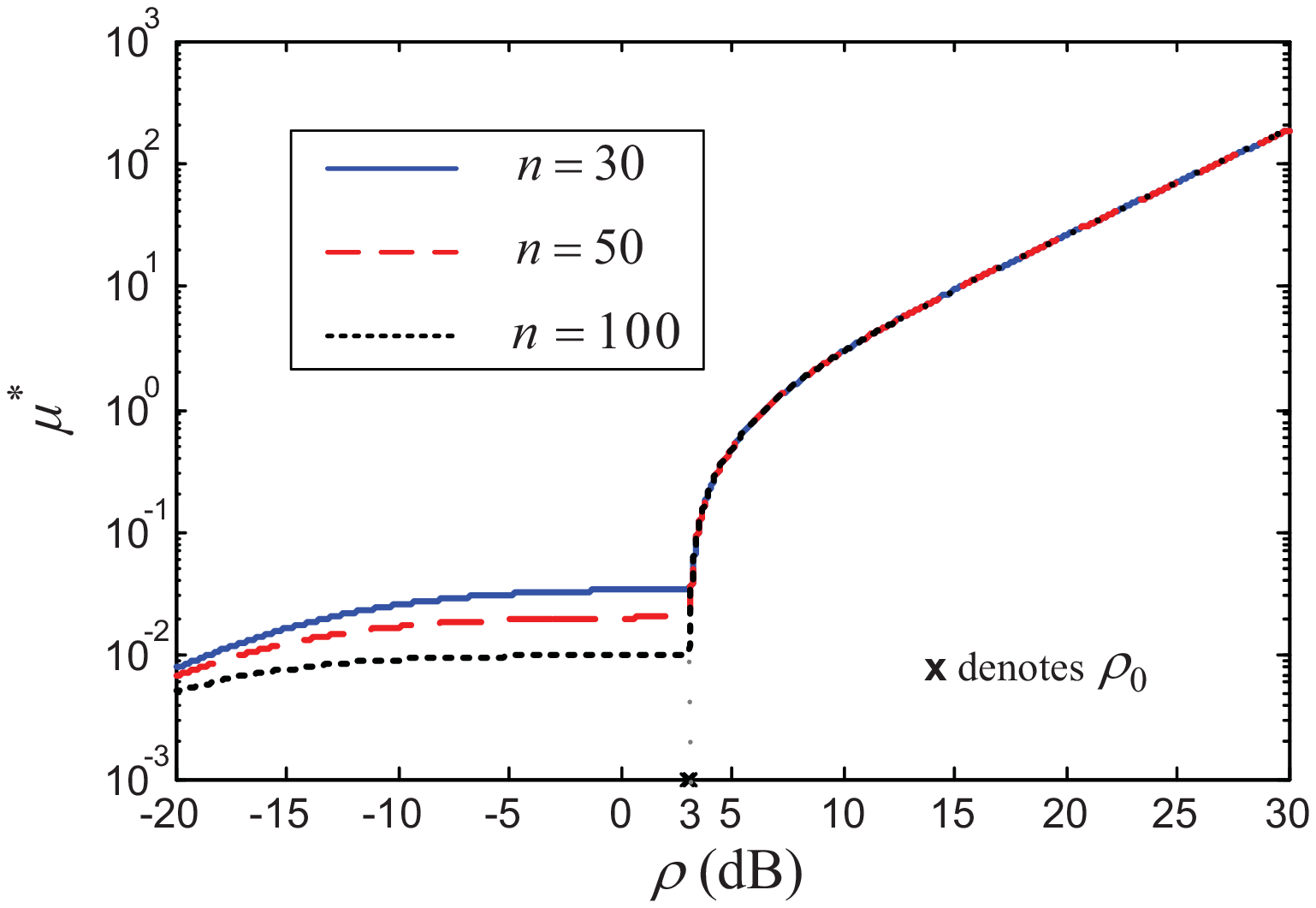}}
\hfil
\subfloat[]{
\label{optimalmaxThroughput}
\includegraphics[width=3.5in,height=2.2in]{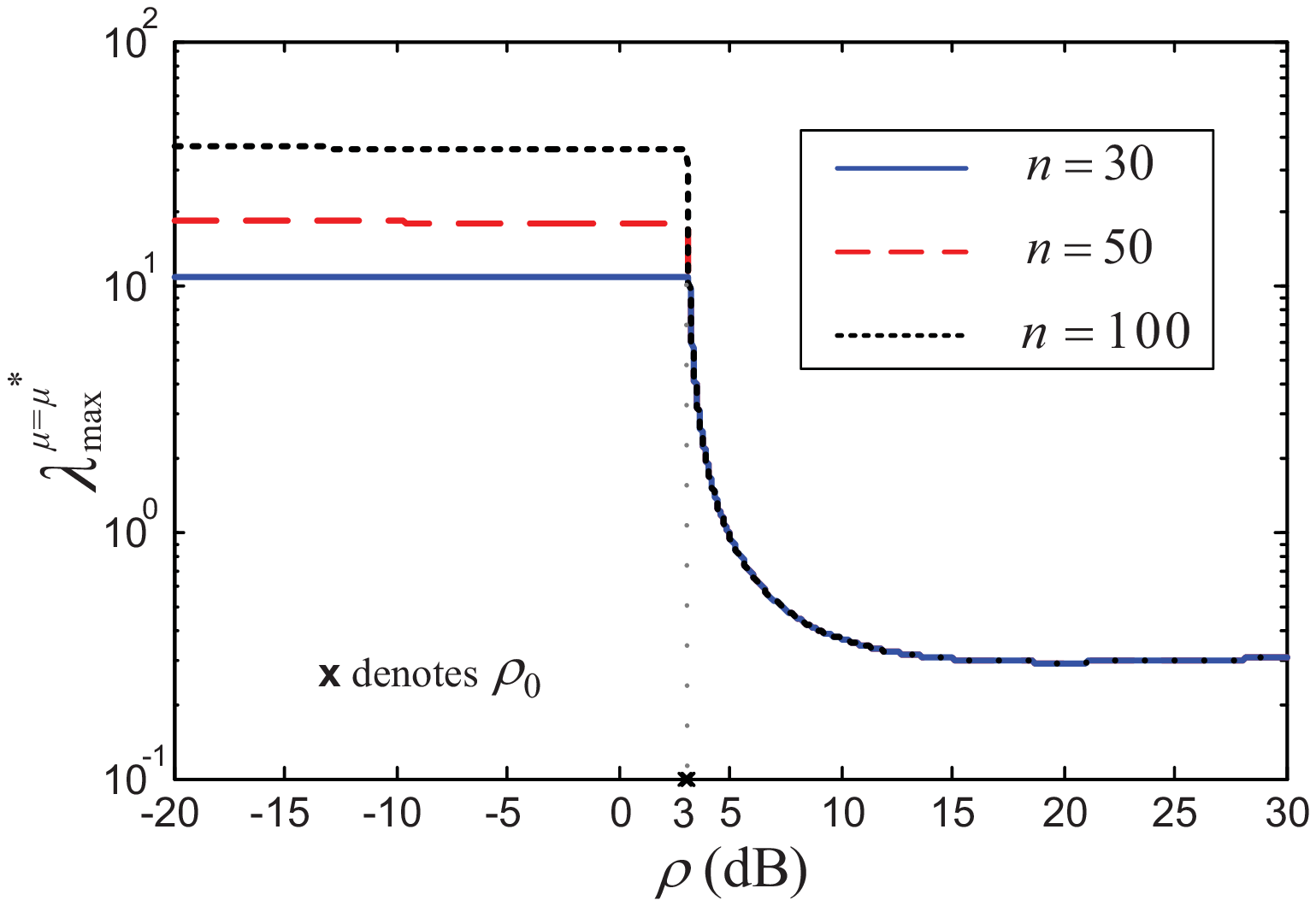}}
\caption{(a) Optimal SINR threshold $\mu^*$ and (b) maximum network throughput $\hat{\lambda}_{\max}^{\mu=\mu^*}$ versus mean received SNR $\rho$. }
\end{figure*}

On the other hand, for given mean received SNR $\rho$, $\hat{\lambda}_{\max}$ monotonically decreases as the SINR threshold $\mu$ increases, as Fig. \ref{maxThroughput-threshold} illustrates. With a lower $\mu$, the receiver can decode more packets among multiple concurrent transmissions, and thus better throughput performance can be achieved. It can be easily shown that multipacket reception is possible when the SINR threshold $\mu$ is sufficiently small. Specifically, for $\mu\geq \tfrac{1}{n-1}$, $\hat{\lambda}_{\max}>1$ if and only if $\tfrac{1}{n-1} \leq\mu<\tfrac{1}{e-1}$ and $\rho>\tfrac{\mu}{\ln \tfrac{\mu+1}{\mu}-1}$. Otherwise, $\hat{\lambda}_{\max}>1$ if and only if $\rho>\tfrac{\mu}{\ln{n}-\tfrac{n\mu}{\mu+1}}$.
As Fig. \ref{maxThroughput-SNR} illustrates, with $n=50$, if the SINR threshold $\mu=0.01<\tfrac{1}{n-1}$, $\hat{\lambda}_{\max}>1$ when the mean received SNR $\rho>-25.3$dB. On the other hand, if $\mu=0.5$, we have $\tfrac{1}{n-1}<\mu<\tfrac{1}{e-1}\approx 0.582$. In this case, $\hat{\lambda}_{\max}>1$ when the mean received SNR $\rho>7$dB.

Note that in spite of the improvement on the maximum network throughput by reducing the SINR threshold $\mu$, the information encoding rate that can be supported for reliable communications, i.e., $R=\log_2(1+\mu)$, is quite low when $\mu$ is small. It is clear that the SINR threshold $\mu$ determines a tradeoff between the network throughput and the information encoding rate. In the next section, we will further study how to maximize the sum rate by properly choosing the SINR threshold $\mu$.

\section{Maximum Sum Rate}\label{Section4}

In this section, we will derive the maximum sum rate and the corresponding optimal SINR threshold as functions of the mean received SNR $\rho$, and discuss their characteristics at the high SNR and lower SNR regions, respectively.

Specifically, it has been demonstrated in Section \ref{Section2} that the sum rate of slotted Aloha networks is determined by the information encoding rate $R$ and the network throughput $\hat\lambda_{out}$. By combining (\ref{mu}) and (\ref{sum rate}), the maximum sum rate can be written as
\begin{equation}\label{maxRate_0}
C=\max_{\mu,\{q_i\}}\left(\hat{\lambda}_{out}\log_2(1{+}\mu)\right){=}\max_{\mu}\left(\log_2(1{+}\mu)\max_{\{q_i\}}\hat{\lambda}_{out}\right).
\end{equation}
Section \ref{Section3} further shows that if backoff parameters $\{q_i\}$ are properly selected, the network throughput is maximized at $\hat\lambda_{\max}$, which is a function of the SINR threshold $\mu$. By combining (\ref{maxRate_0}) and Theorem \ref{Theorem_maxThroughput}, the maximum sum rate can be further written as $C=\max_{\mu>0}f(\mu)$,
where the objective function $f(\mu)$ is given by
\begin{equation}\label{maxRate_f}
f(\mu)=\begin{cases}
\tfrac{\mu+1}{\mu}\exp\left (-1-\tfrac{\mu}{\rho}\right)\log_2(1+\mu) & \text{if\;\;}\mu\geq\tfrac{1}{n-1} \\
n\exp\left(-\tfrac{n\mu}{\mu+1}-\tfrac{\mu}{\rho}\right)\log_2(1+\mu)  & \text{otherwise}.
\end{cases}
\end{equation}
The following theorem presents the maximum sum rate $C$ and the optimal SINR threshold $\mu^*$.

\begin{theorem}\label{Theorem_maximumRate}
For given mean received SNR $\rho\in(0,\infty)$, the maximum sum rate is
\begin{equation}\label{maxRate_2}
C=\begin{cases}
\tfrac{\mu_h^{*}+1}{\mu_h^{*}}\exp\left (-1-\tfrac{\mu_h^{*}}{\rho}\right) \log_2(1+\mu_h^{*}) & \text{if\;\;}\rho\geq \rho_0 \\
n\exp\left(-\tfrac{n\mu_l^{*}}{\mu_l^{*}+1}-\tfrac{\mu_l^{*}}{\rho}\right)\log_2(1+\mu_l^{*})  & \text{otherwise},
\end{cases}
\end{equation}
which is achieved at
\begin{equation}\label{optimal-SINR-threshold}
\mu^*=\begin{cases}
\mu_h^{*} & \text{if\;\;}\rho\geq\rho_0 \\
\mu_l^{*}  & \text{otherwise},
\end{cases}
\end{equation}

\noindent where $\mu_h^{*}$ and $\mu_l^{*}$ are the roots of the following equations:
\begin{equation}\label{optimal-SINR-threshold_root_h}
(\mu+1)^{\tfrac{\mu+1}{\rho}+\tfrac{1}{\mu}}=e,
\end{equation}
and
\begin{equation}\label{optimal-SINR-threshold_root_l}
(\mu+1)^{\tfrac{\mu+1}{\rho}+\tfrac{n}{\mu+1}}=e,
\end{equation}
respectively, and
\begin{equation}\label{threshold_rho}
\rho_0=\tfrac{\tfrac{n}{n-1}\ln{\tfrac{n}{n-1}}}{1-(n-1)\ln{\tfrac{n}{n-1}}}.
\end{equation}
\end{theorem}
\begin{proof}
See Appendix \ref{Proof_Theorem_maximumRate}.
\end{proof}
Note that $\rho_0$ is a monotonic decreasing function of $n\in[2,\infty)$, and $\lim_{n\to\infty}\rho_0=2$. When the number of nodes $n$ is large, $\rho_0$ is close to $3$dB.

\begin{figure*}[!tp]
\centering
\subfloat[]{
\label{maxRate_highSNR}
\includegraphics[width=3.5in,height=2.2in]{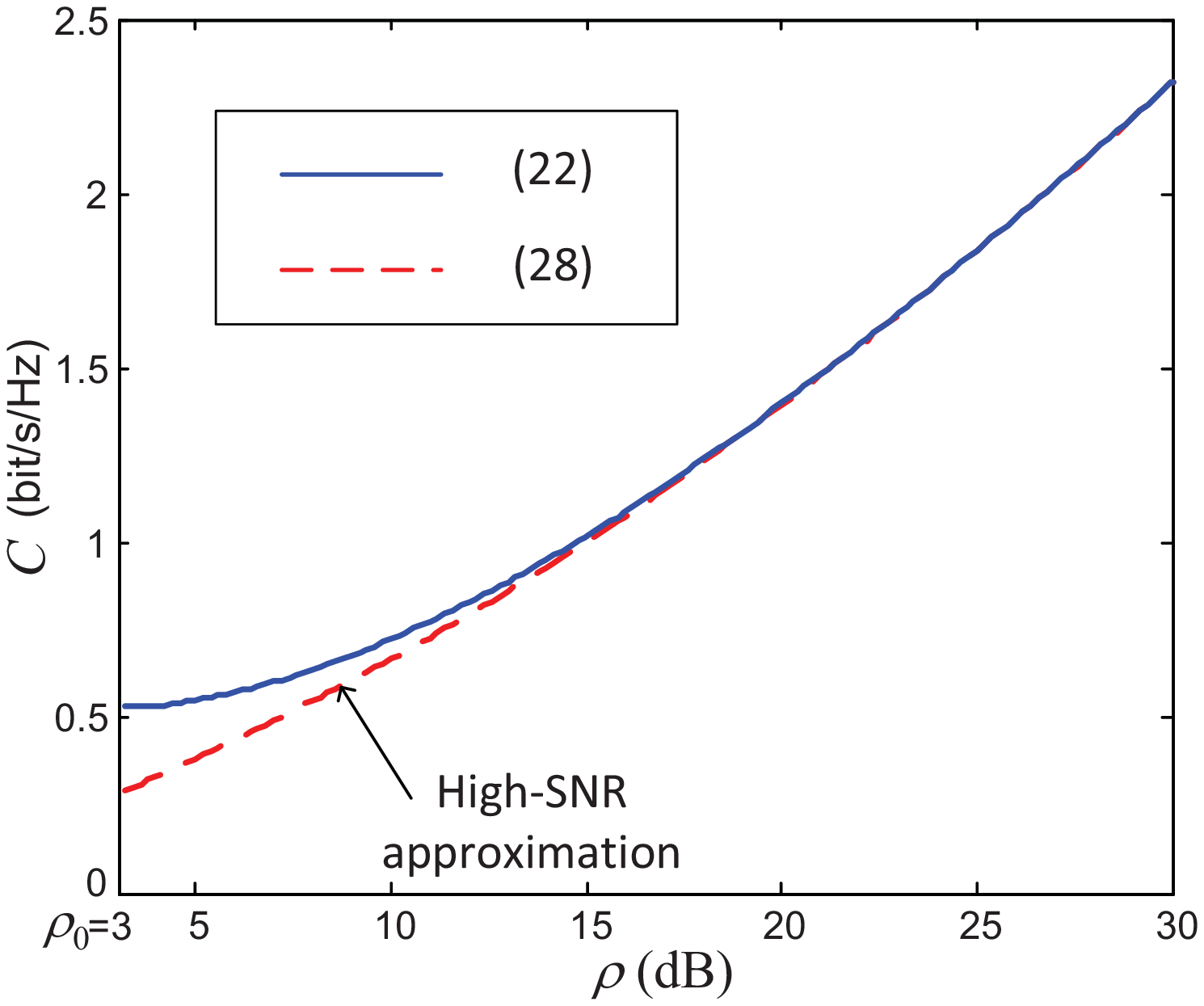}}
\hfil
\subfloat[]{
\label{maxRate_lowSNR}
\includegraphics[width=3.5in,height=2.2in]{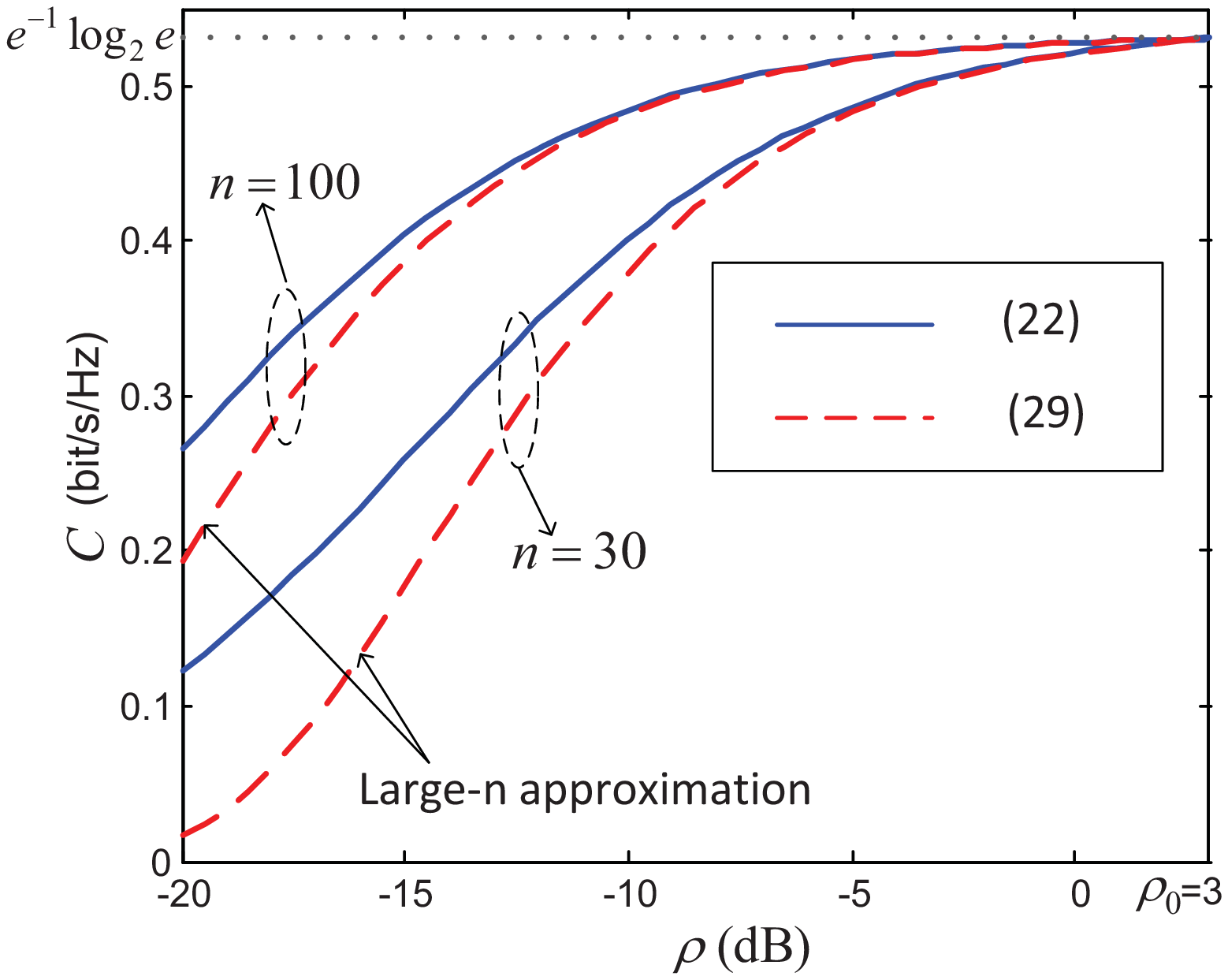}}
\caption{(a) Maximum sum rate $C$ at the high SNR region. (b) Maximum sum rate $C$ at the low SNR region.}
\end{figure*}

\subsection{Optimal SINR Threshold $\mu^*$}\label{Section4-1}
Theorem \ref{Theorem_maximumRate} shows that to achieve the maximum sum rate, the SINR threshold $\mu$ should be carefully selected. Fig. \ref{optimalThreshold} illustrates how the optimal SINR threshold $\mu^*$ varies with the mean received SNR $\rho$. At the low SNR region, i.e., $\rho<\rho_0$, for instance, we can obtain from (\ref{optimal-SINR-threshold}) and (\ref{optimal-SINR-threshold_root_l}) that $\mu_{\rho < \rho_0}^*=\mu_l^{*}\approx e^{-{\mathbb W}_{0}\left(-\tfrac{1}{n}\right)}-1$ for large $n$, where ${\mathbb W}_{0}(z)$ is the principal branch of the Lambert W function \cite{Corless}. In this case, the effect of the mean received SNR $\rho$ becomes negligible, and $\mu_{\rho < \rho_0}^*$ reduces to a monotonic decreasing function of the number of nodes $n$. With a large $n$, $\mu_{\rho < \rho_0}^*\ll 1$, implying that multiple packets can be successfully decoded.

At the high SNR region, we can obtain from (\ref{optimal-SINR-threshold}-\ref{optimal-SINR-threshold_root_h}) that $\mu_{\rho\geq\rho_0}^* =\mu_h^{*}\approx e^{{\mathbb W}_{0}(\rho)}$ for large $\rho$.  As we can see from Fig. \ref{optimalThreshold}, with $\rho\gg1$, the optimal SINR threshold $\mu_{\rho\geq\rho_0}^*$ monotonically increases with the mean received SNR $\rho$.

By combining (\ref{optimal-SINR-threshold}) with Theorem \ref{Theorem_maxThroughput}, we can also obtain the maximum network throughput with $\mu=\mu^*$ as
\begin{equation}\label{eqmtforopmu}
\hat{\lambda}_{\max}^{\mu=\mu^*}=\begin{cases}
\tfrac{\mu_h^{*}+1}{\mu_h^{*}}\exp\left (-1-\tfrac{\mu_h^{*}}{\rho}\right) & \text{if}\;\;\rho\geq\rho_0 \\
n\exp\left(-\tfrac{n\mu_l^{*}}{\mu_l^{*}+1}-\tfrac{\mu_l^{*}}{\rho}\right)   & \text{otherwise}.
\end{cases}
\end{equation}
Fig. \ref{optimalmaxThroughput} illustrates how the maximum network throughput $\hat{\lambda}_{\max}^{\mu=\mu^*}$ varies with the mean received SNR $\rho$. As we can see from Fig. \ref{optimalmaxThroughput}, at the low SNR region, i.e., $\rho<\rho_0$, the effect of $\rho$ is negligible, and $\hat{\lambda}_{\max,\rho<\rho_0}^{\mu=\mu^*}$ becomes a monotonic increasing function of the number of nodes $n$. In this case, the optimal SINR threshold $\mu_{\rho<\rho_0}^*=\mu_{l}^*$ is decreased as $n$ grows, and thus more packets can be successfully decoded, though each at a smaller information encoding rate. For large $n$, we have $\hat{\lambda}_{\max,\rho<\rho_0}^{\mu=\mu^*}\approx ne^{-1}$ according to (\ref{eqmtforopmu}).

At the high SNR region, Fig. \ref{optimalThreshold} has shown that the optimal SINR threshold $\mu_{\rho\geq\rho_0}^* =\mu_h^{*}$ is much larger than $1$, with which at most one packet can be successfully decoded at each time slot. Therefore, the maximum network throughput $\hat{\lambda}_{\max,\rho\geq\rho_0}^{\mu=\mu^*}$ quickly drops below $1$, and eventually approaches $e^{-1}$ as $\rho\to\infty$.

\subsection{Maximum Sum Rate $C$ at High SNR Region}\label{Section4-2}

Similar to Section \ref{Section4-1}, let us take a closer look at the maximum sum rate $C$ at different SNR regions.


With $\rho\geq\rho_0$, it has been shown in Section \ref{Section4-1} that the optimal SINR threshold $\mu_{\rho\geq\rho_0}^* =\mu_h^{*}\approx e^{{\mathbb W}_{0}(\rho)}$ for large $\rho$. The maximum sum rate in this case can be then approximated by
\begin{equation}\label{CForHighR}
C_{\rho\geq\rho_0}\approx \left(1+e^{-{\mathbb W}_{0}(\rho)}\right)\exp\left (-1-\tfrac{e^{{\mathbb W}_{0}(\rho)}}{\rho}\right)\log_2(1+e^{{\mathbb W}_{0}(\rho)}),
\end{equation}
for $\rho\gg 1$.
As Fig. \ref{maxRate_highSNR} shows, the approximation (\ref{CForHighR}) works well when the mean received SNR $\rho$ is large, i.e., $\rho\geq15$dB. Moreover, a logarithmic increase of the maximum sum rate $C$ can be observed at the high SNR region. The following corollary presents the high-SNR slope of $C$.
\begin{corollary}\label{Corollary_highSNR}
$\lim_{\rho\rightarrow\infty}\tfrac{C}{\log_2\rho}=e^{-1}$.
\end{corollary}
\begin{proof}
See Appendix \ref{Proof_Corollary_highSNR}.
\end{proof}

Recall that the high-SNR slope of the ergodic sum capacity of MAC is $1$ when single-antenna is employed at both the transmitters and the receiver. To achieve the ergodic sum capacity, however, a joint decoding of all received signals is required and the codewords should span multiple fading states. With the capture model, in contrast, each node's packet is decoded independently by treating others' as background noise at each time slot. When the mean received SNR is high, at most one packet can be successfully decoded each time due to a large SINR threshold $\mu^*\gg 1$. Corollary \ref{Corollary_highSNR} shows that with the simplified receiver, the high-SNR slope of the maximum sum rate of slotted Aloha networks is significantly lower than that of the sum capacity.

\begin{figure*}[!tp]
\centering
\subfloat[]{
\label{pA_K}
\includegraphics[width=3.5in,height=2.2in]{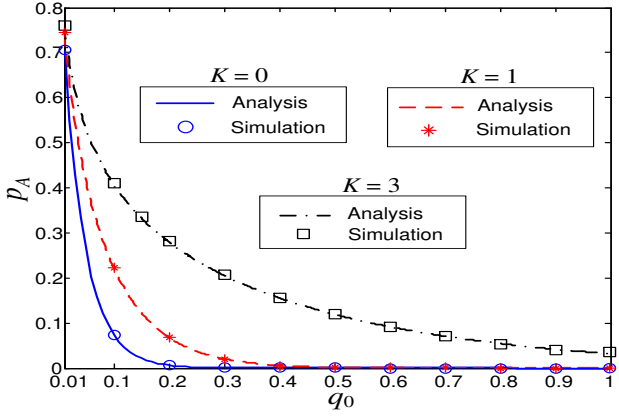}}
\hfil
\subfloat[]{
\label{pA_n}
\includegraphics[width=3.5in,height=2.2in]{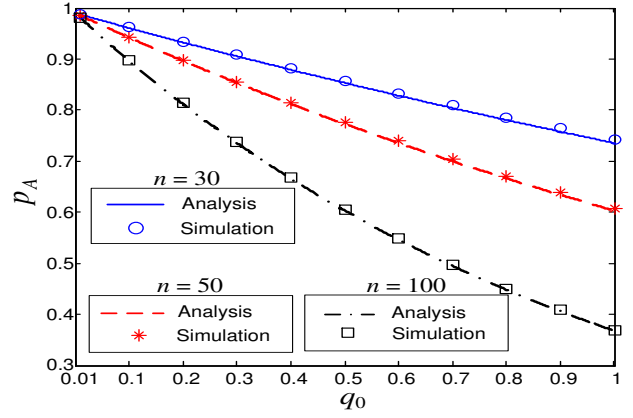}}
\caption{Steady-state point $p_A$ versus initial transmission probability $q_0$. (a) $n=50$. $\mu=1$ and $\rho=10$dB. (b) $K=0$. $\mu=0.01$ and $\rho=0$dB.}
\label{sim_pA}
\end{figure*}

\begin{figure*}[!tp]
\centering
\subfloat[]{
\label{simThroughput_K}
\includegraphics[width=3.5in,height=2.2in]{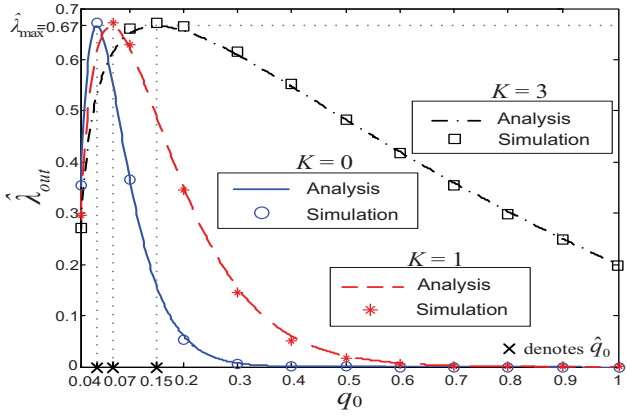}}
\hfil
\subfloat[]{
\label{simThroughput_n}
\includegraphics[width=3.5in,height=2.2in]{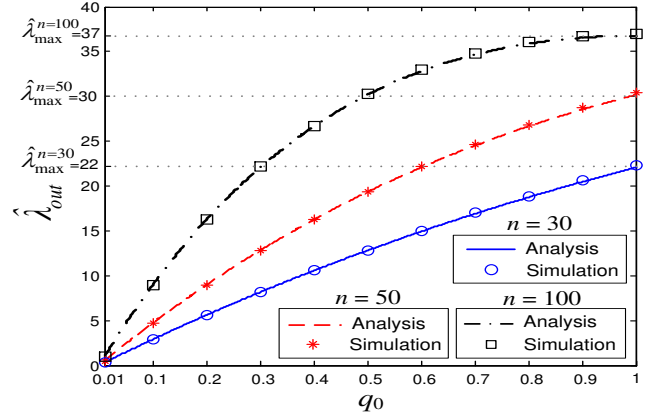}}
\caption{Network throughput $\hat{\lambda}_{out}$ versus initial transmission probability $q_0$. (a) $n=50$. $\mu=1$ and $\rho=10$dB. (b) $K=0$. $\mu=0.01$ and $\rho=0$dB.}
\label{sim_throughput}
\end{figure*}

\subsection{Maximum Sum Rate $C$ at Low SNR Region}\label{Section4-3}

For $\rho<\rho_0$, it has been shown in Section \ref{Section4-1} that the optimal SINR threshold $\mu_{\rho<\rho_0}^*=\mu_{l}^{*}\approx e^{-{\mathbb W}_{0}\left(-\tfrac{1}{n}\right)}-1$ for large $n$. The corresponding maximum sum rate can be then approximated by
\begin{align}\label{lowSNRlimit}
C_{\rho<\rho_0}&\approx -n{\mathbb W}_{0}\left(-\tfrac{1}{n}\right) \notag \\
&\cdot \exp\left(-n\left(1-e^{{\mathbb W}_{0}\left(-\tfrac{1}{n}\right)}\right)-\tfrac{e^{-{\mathbb W}_{0}\left(-\tfrac{1}{n}\right)}-1}{\rho}\right) \log_2 e,
\end{align}
for $n\gg 1$.
As we can see from Fig. \ref{maxRate_lowSNR}, the approximation (\ref{lowSNRlimit}) works well when the number of nodes $n$ is large. The following corollary further presents the limiting maximum sum rate as $n\to\infty$ at the low SNR region.

\begin{corollary}\label{Corollary_lowSNR}
$\lim_{n\rightarrow\infty}C_{\rho<\rho_0}=e^{-1}{\log_2 e}$.
\end{corollary}
\begin{proof}
See Appendix \ref{Proof_Corollary_lowSNR}.
\end{proof}

Note that it has been shown in Section \ref{Section4-1} that with $\rho<\rho_0$, the maximum network throughput $\hat{\lambda}_{\max}^{\mu=\mu^*}\approx ne^{-1}$, which grows with the number of nodes $n$ unboundedly. Although more packets can be successfully decoded, the information carried by each packet decreases as $n$ increases due to a diminishing information encoding rate, i.e., $R=\log_2(1+\mu^*_{\rho<\rho_0})\approx \tfrac{1}{n}\log_2 e$ for large $n$. Therefore, as the number of nodes $n\to\infty$, the maximum sum rate reaches a limit that is independent of the mean received SNR, as Corollary \ref{Corollary_lowSNR} indicates. It is in sharp contrast to the ergodic sum capacity of MAC which linearly increases with $n$ and $\rho$ at the low SNR region.

\section{Simulation Results}\label{Section5}

In this section, simulation results will be presented to verify the preceding analysis in Sections \ref{Section3} and \ref{Section4}. In particular, we consider a saturated slotted Aloha network with Binary Exponential Backoff (BEB), i.e., the transmission probabilities of each node are given by the geometric series $q_i=q_0\cdot\tfrac{1}{2^i}$, $i=0, \ldots, K$. The simulation setting is the same as the system model and thus we omit the details here.

Section \ref{Section3-2} has shown that the network operates at the steady-state point $p_A$, which is closely determined by the number of nodes $n$ and the backoff parameters $\{q_i\}$. The expression of $p_A$ is given in $\eqref{Probability-of-success_4}$ and verified by the simulation results presented in Fig. \ref{sim_pA}.\footnote{In simulations, the steady-state probability of successful transmission of HOL packets $p_A$ is obtained by calculating the ratio of the number of successful transmissions to the total number of attempts of HOL packets over a long time period, i.e., $10^8$ time slots.}

\begin{figure}[htb]
\centering
\includegraphics[width=3.2in,height=2in]{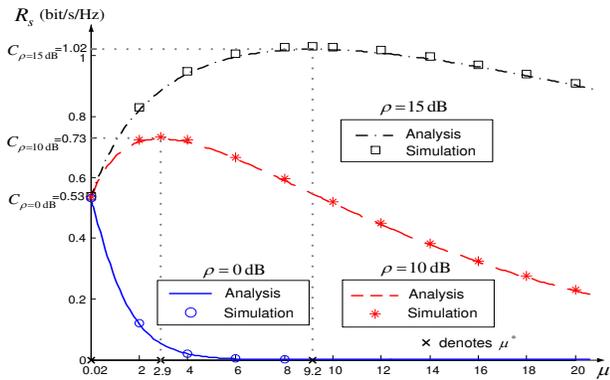}\
\caption{Sum rate $R_s$ versus SINR threshold $\mu$ under different values of mean received SNR $\rho$. $n=50$. $K=0$ and $q_0=q_0^{*}$.}
\label{simMaxRate}
\end{figure}

Fig. \ref{sim_throughput} illustrates the corresponding network throughput performance. The network throughput $\hat{\lambda}_{out}$ has been derived as a function of $p_A$ in (\ref{eq15}) in Section \ref{Section3-3}, which varies with the backoff parameters. As we can see from Fig. \ref{sim_throughput}, the network throughput performance is sensitive to the setting of the initial transmission probability $q_0$. According to Theorem \ref{Theorem_maxThroughput}, when the SINR threshold $\mu\geq\tfrac{1}{n-1}$, the maximum network throughput $\hat{\lambda}_{\max}$ is achieved when $q_0$ is set to be $\hat{q}_0$. Otherwise, $\hat{\lambda}_{\max}$ is achieved with $q_i{=}1$, $i=0,\ldots,K$. The expressions of $\hat{\lambda}_{\max}$ and the corresponding optimal backoff parameters $q_i^{*}$ are given in \eqref{maxThroughput_all} and \eqref{maxThroughput_q}, respectively, and verified by the simulation results presented in Fig. \ref{sim_throughput}.

It is further demonstrated in Section \ref{Section4} that as both the maximum network throughput and the information encoding rate depend on the SINR threshold $\mu$, the sum rate can be maximized by optimally choosing $\mu$. We can clearly observe from Fig. \ref{simMaxRate} that the sum rate performance is sensitive to the SINR threshold $\mu$ especially when the mean received SNR $\rho$ is small. To achieve the maximum sum rate, $\mu$ should be properly set according to $\rho$. The expressions of the optimal SINR threshold $\mu^*$ and the maximum sum rate $C$ are given in Theorem \ref{Theorem_maximumRate}, and verified by the simulation results presented in Fig. \ref{simMaxRate}.


\section{Discussions}\label{Section6}

So far we have shown that to optimize the sum rate performance of slotted Aloha networks, the SINR threshold $\mu$ and backoff parameters $\{q_i\}$ should be properly set according to the mean received SNR $\rho$, and the maximum sum rate logarithmically increases with $\rho$ with the high-SNR slope of $e^{-1}$. In this section, we will further discuss how the performance is affected by key factors such as backoff and power control.

\subsection{Effect of Adaptive Backoff}\label{Section5-1}
\newcounter{TempEqCnt}
\setcounter{TempEqCnt}{\value{equation}}
\setcounter{equation}{34}
\begin{figure*}[ht]
\begin{equation}\label{i_m-concurrent-in-group_m}
\text{Pr\{$i_m$ concurrent transmissions in Group $m$\}}=\begin{cases}
\binom{n_m}{i_m}\left(1-\pi_T^{(m)}/p^{(m)}\right)^{n_m-i_m}\cdot \left(\pi_T^{(m)}/p^{(m)}\right)^{i_m} & m\neq l\\
\binom{n_l-1}{i_l}\left(1-\pi_T^{(l)}/p^{(l)}\right)^{n_l-1-i_l}\cdot \left(\pi_T^{(l)}/p^{(l)}\right)^{i_l} & m=l.
\end{cases}
\end{equation}
\hrulefill
\end{figure*}
\setcounter{equation}{\value{TempEqCnt}}

Backoff is a key component of random-access networks. It has been shown in Sections \ref{Section3} and \ref{Section4} that to achieve the maximum sum rate, backoff parameters, i.e., the transmission probabilities $\{q_i\}$ of nodes, should be adaptively tuned according to the number of nodes $n$ and the mean received SNR $\rho$.\footnote{Note that for practical random-access networks, the backoff parameters can be updated through the feedback from the common receiver. In IEEE 802.11 networks, for instance, as each node associates with the access-point (AP) upon joining the network, the AP can count the number of nodes through the MAC header of the frame sent by each node, calculate the optimal backoff parameters, and broadcast them in the beacon frame periodically. Each node can then update its backoff parameters according to the received beacon frame. Such a feedback-based update process can also be implemented in cellular systems where the base-station serves as the common receiver in each cell.} In many studies, however, nodes are supposed to transmit their packets with a fixed probability \cite{Namislo,Goodman,Habbab,Rasool,Dua}.
To see how the rate performance of slotted Aloha deteriorates without adaptive backoff, let us assume that each node transmits its packet with a constant probability $q$ at each time slot, i.e., $q_i=q$, $i=0, \ldots, K$. In this case, the network steady-state point in saturated conditions can be obtained from \eqref{Probability-of-success_4} as $p_A^{q_i=q}=\exp\left(-\tfrac{\mu}{\rho}-\tfrac{nq\mu}{\mu+1}\right)$,
and the corresponding network throughput is $\hat{\lambda}_{out}^{q_i=q}=nq \exp\left(-\tfrac{\mu}{\rho}-\tfrac{nq\mu}{\mu+1}\right)$,
according to \eqref{eq15}. The sum rate can be then written as $R_s^{q_i=q}=nq\exp\left(-\tfrac{\mu}{\rho}-\tfrac{nq\mu}{\mu+1}\right)\cdot \log_2(1+\mu)$,
which is an increasing function of the mean received SNR $\rho$.

As $\rho\to\infty$, it can be easily obtained that $\tilde{R}_s^{q_i=q}=\lim_{\rho\to\infty}R_s^{q_i=q}=nq\exp\left(-\tfrac{nq\mu}{\mu+1}\right)\cdot \log_2(1+\mu)$, with the maximum
\begin{align}\label{maxsum-rate_no-backoff}
\max_\mu\tilde{R}_s^{q_i=q}&=nq\exp\left(-nq\left(1-e^{{\mathbb W}_{0}\left(-\tfrac{1}{nq}\right)}\right)\right) \notag \\
&\cdot \log_2e^{-{\mathbb W}_{0}\left(-\tfrac{1}{nq}\right)},
\end{align}
which is achieved at
\begin{equation}\label{optimal-threshold_no-backoff}
\mu^{*,q_i=q}=e^{-{\mathbb W}_{0}\left(-\tfrac{1}{nq}\right)}-1.
\end{equation}
Eq. \eqref{optimal-threshold_no-backoff} shows that the optimal SINR threshold $\mu^{*,q_i=q}$ monotonically decreases as the number of nodes $n$ grows. For large $n\gg 1$, it can be obtained from (\ref{maxsum-rate_no-backoff}-\ref{optimal-threshold_no-backoff}) that $\mu^{*,q_i=q}\approx \tfrac{1}{nq}$, and 
\begin{equation}\label{limmaxsum-rate_no-backoff}
\max_\mu \tilde{R}_{s}^{q_i=q}\mathop{\approx }\limits^{n\gg 1} e^{-1}{\log_2 e}.
\end{equation}
Recall that it has been shown in Section \ref{Section4-2} that the maximum sum rate increases with the mean received SNR $\rho$ unboundedly. Here (\ref{limmaxsum-rate_no-backoff}) indicates that with a constant transmission probability, the sum rate converges to a limit that is much lower than $1$ as $\rho\to\infty$. It corroborates that adaptive backoff is indispensable for random-access networks.

It is interesting to note that when $q=1$, all the nodes persistently transmit their packets, and the slotted Aloha network reduces to a typical MAC. It is well known that for an $n$-user AWGN MAC, if the capture model is adopted at the receiver side
and all the users have equal received power, the sum rate approaches $\log_2 e$ as $n\to\infty$ \cite{Tse}. Here we can see from (\ref{limmaxsum-rate_no-backoff}) that an additional factor of $e^{-1}$ is introduced, which is mainly attributed to the effect of channel fading.\footnote{Note that in this paper, each codeword is assumed to last for one channel coherence time period. Without coding over different fading states, the channel fluctuations cannot be averaged out, thus leading to a significant rate loss compared to the AWGN case.}

\subsection{Effect of Power Control}\label{Section5-3}

So far we have focused on a homogeneous slotted Aloha network where all the nodes have the same mean received SNR $\rho$. In this section, the analysis will be extended to the heterogeneous case, where nodes in the same group have an identical mean received SNR but SNRs differ from group to group.

Specifically, assume that $n$ nodes are divided into $M$ groups. Group $m$ has $n_m$ nodes, and each node in Group $m$ has the mean received SNR $\rho_m$, $m=1,\ldots, M$. For HOL packet $j$, let $\mathcal{S}_j$ denote the set of nodes that have concurrent transmissions. It can be successfully decoded at the receiver if and only if its received SINR is above the SINR threshold $\mu$, i.e., $\tfrac{P_j}{\sum_{k\in \mathcal{S}_j}P_k+\sigma^2}\geq \mu$, where $P_k$ denotes the received power of node $k$'s packet. Suppose that $\mathcal{S}_j=\bigcup_{m=1,\ldots,M} \mathcal{S}_j^{m}$, where $\mathcal{S}_j^{m}$ denotes the set of nodes which have concurrent transmissions in Group $m$, and $|\mathcal{S}_j^{m}|=i_m$, $m=1,\ldots,M$. The steady-state probability of successful transmission of HOL packet $j$ given that there are $\{i_m\}_{m=1,\ldots,M}$ concurrent transmissions, $r_{\{i_m\}}^j$, can be then written as $r_{\{i_m\}}^j=\textrm{Pr}\left\{\tfrac{|h_j|^2}{\sum_{m=1}^{M}\sum_{k \in \mathcal{S}_j^{m}  } |h_k|^2\cdot \tfrac{\rho_m}{\rho_j}+\tfrac{1}{\rho_j}}\geq\mu\right\}$.
It can be easily shown that with $h_k\sim \mathcal{CN}(0,1)$, $r_{\{i_m\}}^j$ is given by
\begin{equation}\label{conditional probability2}
r_{\{i_m\}}^j=\tfrac{\exp\left(-\tfrac{\mu}{\rho_j}\right)}{\prod_{m=1}^M \left(1+\tfrac{\rho_m}{\rho_j}\mu\right)^{i_m}}.
\end{equation}
It can be seen from (\ref{conditional probability2}) that $r_{\{i_m\}}^j$ is determined by the mean received SNR $\rho_j$ of HOL packet $j$. Suppose that HOL packet $j$ belongs to Group $l\in\{1,\cdots,M\}$. As nodes in the same group have an identical mean received SNR, the superscript $j$ can be replaced by its group index $l$. The steady-state probability of successful transmission of HOL packet $j$ in Group $l\in\{1,\cdots,M\}$ can then be written as
\begin{align}\label{pj}
p^{(l)}&=\sum_{i_1=0}^{n_1}\cdots \sum_{i_l=0}^{n_l-1}\cdots \sum_{i_M=0}^{n_M} r_{\{i_m\}}^l \notag \\
&\cdot \prod_{m=1}^{M}\text{Pr\{$i_m$ concurrent transmissions in Group $m$\}}.
\end{align}
For each node in Group $m$, the probability that it is busy with the HOL packet requesting transmission in saturated conditions is given by $\pi_T^{(m)}q_0+\sum_{i=0}^K \pi_i^{(m)} q_i$, which is equal to $\pi_T^{(m)}/p^{(m)}$ according to (\ref{piK}). Therefore, we have \eqref{i_m-concurrent-in-group_m}, which is shown at the top of this page.

By combining (\ref{conditional probability2}-\ref{i_m-concurrent-in-group_m}), the steady-state probability of successful transmission of HOL packet $j$ in Group $l\in\{1,\cdots,M\}$ can be obtained as
\setcounter{equation}{35}
\begin{align}\label{pj_expression1}
p^{(l)}&{=}\exp \left({-}\tfrac{\mu}{\rho_l}\right)\cdot \left(1{-}\tfrac{\mu}{\mu{+}1}\cdot \tfrac{\pi_T^{(l)}}{p^{(l)}}\right)^{n_l{-}1} \notag\\
&\cdot\prod_{m{=}1,m\neq l}^M \left(1{-}\tfrac{\mu}{\mu{+}\rho_l/\rho_m}{\cdot} \tfrac{\pi_T^{(m)}}{p^{(m)}}\right)^{n_m} \notag\\
&\stackrel{\text{for large}\;  n_1,\ldots ,n_M}{\approx} \exp\left(-\tfrac{\mu}{\rho_l}-\sum_{m=1}^M \tfrac{n_m \mu}{\mu+\rho_l/\rho_m}\cdot \tfrac{\pi_T^{(m)}}{p^{(m)}} \right).
\end{align}
Finally, by substituting (\ref{pi0}) into (\ref{pj_expression1}), we have
\begin{align}\label{pj_expression}
p^{(l)}&{=}\exp\left({-}\tfrac{\mu}{\rho_l}{-}\sum_{m=1}^M \tfrac{n_m \mu}{\mu{+}\rho_l/\rho_m}\right. \notag \\
&\left.{\cdot} \tfrac{1}{\sum_{i{=}0}^{K{-}1}\tfrac{p^{(m)}\left(1{-}p^{(m)}\right)^i}{q_i}{+}\tfrac{\left(1{-}p^{(m)}\right)^K}{q_K}} \right),
\end{align}
$l\in\{1,\cdots,M\}$. We can see from \eqref{pj_expression} that in the heterogeneous case, HOL packets in different groups have distinct steady-state probabilities of successful transmission. With $M$ groups, $M$ non-zero roots $\{p_A^{(m)}\}_{m=1,\ldots, M}$ can be obtained by jointly solving $M$ fixed-point equations given in \eqref{pj_expression}. Note that nodes in the same group have the same steady-state probability of successful transmission and thus the same throughput performance. For each node in Group $m$, $m=1,\ldots,M$, the node throughput can be obtained from (\ref{pi0}) as
\begin{equation}\label{Throughput_No-power-control}
{\lambda}_{out}^{(m)}=\pi_T^{(m)}= \tfrac{1}{\sum_{i=0}^{K-1}\tfrac{\left(1-p_A^{(m)}\right)^i}{q_i}+\tfrac{\left(1-p_A^{(m)}\right)^K}{p_A^{(m)}q_K}},
\end{equation}
and the network throughput is $\hat{\lambda}_{out}=\sum_{m=1}^M n_m {\lambda}_{out}^{(m)}$.

To illustrate the above results, let us focus on the two-group case and assume that the cutoff phase $K=0$. The steady-state probabilities of successful transmission of HOL packets in Group $1$ and Group $2$ can be obtained from \eqref{pj_expression} as
\begin{equation*}
p_A^{(1)}= \exp \left(-\tfrac{\mu}{\rho_1}-\tfrac{n_1\mu q_0}{\mu+1}-\tfrac{n_2\mu q_0}{\mu+\rho_1/\rho_2}\right),
\end{equation*}
\begin{equation}\label{pA-group1}
p_A^{(2)}= \exp\left(-\tfrac{\mu}{\rho_2}-\tfrac{n_1 \mu q_0}{\mu+\rho_2/\rho_1}-\tfrac{n_2\mu q_0}{\mu+1}\right).
\end{equation}
%
By combining (\ref{pA-group1}) with \eqref{Throughput_No-power-control}, the node throughput can be obtained as
\begin{equation*}
{\lambda}_{out}^{(1)}=q_0 \exp \left(-\tfrac{\mu}{\rho_1}-\tfrac{n_1\mu q_0}{\mu+1}-\tfrac{n_2\mu q_0}{\mu+\rho_1/\rho_2}\right),
\end{equation*}
\begin{equation}\label{Throughput_node1}
{\lambda}_{out}^{(2)}=q_0 \exp\left(-\tfrac{\mu}{\rho_2}-\tfrac{n_1 \mu q_0}{\mu+\rho_2/\rho_1}-\tfrac{n_2\mu q_0}{\mu+1}\right).
\end{equation}

\noindent Eq. (\ref{Throughput_node1}) shows that the throughput performance is closely determined by the mean received SNRs. If the two groups have equal mean received SNRs $\rho_1{=}\rho_2{=}\rho$, for instance, we can see from (\ref{pA-group1}) that all the HOL packets have the same steady-state probability of successful transmission, i.e., $p_A^{(1)}{=}p_A^{(2)}$. The node throughput can be obtained from (\ref{Throughput_node1}) as ${\lambda}_{out}^{(1)}{=}{\lambda}_{out}^{(2)}{=}q_0\exp\left({-}\tfrac{\mu}{\rho}{-}\tfrac{(n_1+n_2)\mu q_0}{\mu+1}\right)$. In this case, each node has an equal probability of accessing the channel, thus achieving the same throughput performance. As the difference between $\rho_1$ and $\rho_2$ grows, nevertheless, the node throughput performance becomes increasingly polarized. We can see from (\ref{pA-group1}) that with $\rho_1 {\gg} \rho_2$, $p_A^{(1)}{\gg} p_A^{(2)}$, which indicates that much more packets from Group $1$ can be successfully received than Group $2$. The throughput performance of nodes in Group $1$ is then much better than that in Group $2$, i.e., ${\lambda}_{out}^{(1)}{\gg}{\lambda}_{out}^{(2)}$ according to (\ref{Throughput_node1}), implying serious unfairness among nodes.

\begin{figure}[!tp]
\centering
\includegraphics[width=3.2in,height=2in]{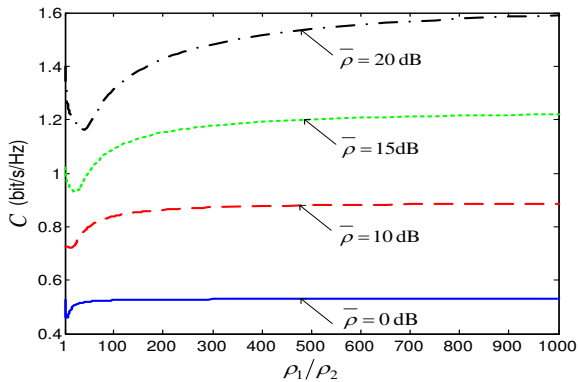}\
\caption{Maximum sum rate versus $\rho_1/\rho_2$ for a two-group slotted Aloha network. $n_1=n_2=25$. $K=0$.}
\label{maxRate_powercontrol}
\end{figure}

As the maximum network throughput $\hat{\lambda}_{\max}=\max_{q_0} \hat{\lambda}_{out}$ does not have an explicit expression in general, we can only numerically calculate the maximum sum rate $C=\max_{\mu} \hat{\lambda}_{\max}\cdot \log_2(1+\mu)$. Fig. \ref{maxRate_powercontrol} illustrates how the maximum sum rate $C$ varies with the ratio of $\rho_1$ and $\rho_2$ by fixing the mean SNR of nodes $\bar{\rho}=\tfrac{\sum_{m=1}^2 n_m\rho_m}{\sum_{m=1}^2 n_m}$ to $0$dB, $10$dB, $15$dB and $20$dB. It is interesting to note from Fig. \ref{maxRate_powercontrol} that with a large SNR ratio $\rho_1/\rho_2\gg 1$, the maximum sum rate is higher than that with $\rho_1/\rho_2=1$, which suggests that despite serious unfairness, the sum rate performance may be improved by introducing a large SNR difference among nodes. Intuitively, the channel efficiency is maximized by allocating all the resources to the strongest node(s). Here we can see that even without a central controller for resource allocation, the fundamental tradeoff between efficiency and fairness still holds true for random-access networks.


The tradeoff nevertheless becomes less significant when the network operates at the low SNR region. It can be observed from Fig. \ref{maxRate_powercontrol} that with $\bar{\rho}=0$dB, the maximum sum rate is insensitive to the SNR ratio. It indicates that power control is desirable in this case, with which the fairness performance can be improved without sacrificing the sum rate.

\section{Conclusion}\label{Section6}

In this paper, the unified analytical framework proposed in \cite{Dai_Aloha,Dai_CSMA} is extended to incorporate the capture model. By assuming that the received SNRs of nodes' packets are exponentially distributed with the same mean received SNR $\rho$ in a saturated slotted Aloha network, explicit expressions of the maximum network throughput and the corresponding optimal backoff parameters are obtained, based on which the maximum sum rate is derived by optimizing the SINR threshold $\mu$. The analysis shows that with a low SNR, the maximum sum rate linearly increases with the number of nodes $n$, and approaches $e^{-1}\log_2 e$ as $n\to\infty$. At the high SNR region, a logarithmic growth of the maximum sum rate is observed as $\rho$ increases, with the high-SNR slope of $e^{-1}$. Effects of key factors, including backoff and power control, on the sum rate performance are also studied.

The analysis sheds important light on the practical network design. For instance, it is demonstrated that to achieve the maximum sum rate, the transmission probabilities of nodes should be adaptively tuned according to the network size and the mean received SNR $\rho$. With a fixed transmission probability, the sum rate may significantly deteriorate, and converges to a limit that is much lower than 1 as $\rho\to\infty$. Moreover, the throughput performance of each node is found to be closely dependent on its mean received SNR. Although a large SNR difference among nodes may be beneficial to the sum rate performance, it introduces serious unfairness. A uniform mean received SNR is shown to be crucial for achieving a good balance between fairness and sum rate when the network operates at the low SNR region.

Note that the proposed analytical framework can be applied to both saturated and unsaturated networks. In this paper, we focus on the saturated conditions where the network throughput is pushed to the limit, yet the mean queueing delay is infinite and the network could be unstable. It is of great practical significance to further study the maximum sum rate of slotted Aloha under certain system constraints such as stability or delay requirements. Moreover, the analysis is based on the capture model, which is essentially a single-user detector. Performance gains on the maximum sum rate and network throughput can be expected if multiuser detectors, such as SIC, are adopted. It is therefore important to further extend the analysis to incorporate more advanced receiver structures. Finally, a key assumption throughout the paper is that the nodes are unaware of the instantaneous realizations of the small-scale fading, and they encode their packets independently at the same rate.
How to characterize the maximum sum rate with CSI at the transmitter side is another interesting and challenging issue, which deserves much attention in the future study.

\appendices
\section{Proof of Theorem \ref{Theorem_pA}} \label{Proof_Theorem_pA}
\begin{IEEEproof}
The right-hand side of \eqref{Probability-of-success_4} can be written as $h(p)=\exp \left(-\tfrac{\mu}{\rho}-\tfrac{n\mu}{\mu+1}\cdot \tfrac{1}{g(p)} \right)$,
where $g(p)=\sum_{i=0}^{K-1}\tfrac{p\left(1-p\right)^i}{q_i}+\tfrac{\left(1-p\right)^K}{q_K}$. Define $\tilde q_i=1/q_i$, for $0\leq i\leq K-1$, and $\tilde q_i=1/q_K$ for $i\geq K$. $g(p)$ can be then written as $g(p)=\sum_{i=0}^\infty p(1-p)^i \tilde q_i = E_X \left[\tilde q_X\right]$,
where $X$ is a geometric random variable with parameter $p$.

Suppose that $0<p_1<p_2\leq1$. Let $X_1$ and $X_2$ denote geometric random variables with parameters $p_1$ and $p_2$, respectively. Then we have $X_1 \geq_{st} X_2$ \cite{Shaked}.\footnote{$X_1 \geq_{st} X_2$ denotes that a random variable $X_1$ is larger than a random variable $X_2$ in the usual stochastic order, i.e., $\text{Pr}(X_1>x)\ge \text{Pr}(X_2>x)$ for all $x\in(-\infty, \infty)$.} As $\{q_i\}$ is a monotonic non-increasing sequence, we have $\tilde q_{X_1} \geq_{st} \tilde q_{X_2}$. We can then conclude that $g(p_1) \geq g(p_2)$. Therefore, $g(p)$ is a monotonic non-increasing function with respect to $p$, which indicates that $h(p)$ is a monotonic non-increasing function. Moreover, as $\lim_{p\rightarrow 0}h(p)=\exp \left(-\tfrac{\mu}{\rho}-\tfrac{n\mu}{\mu+1}\cdot q_K \right)>0$ and $\lim_{p\rightarrow 1}h(p)=\exp \left(-\tfrac{\mu}{\rho}-\tfrac{n\mu}{\mu+1}\cdot q_0 \right)<1$, we can then conclude that \eqref{Probability-of-success_4} has a single non-zero root if $\{q_i\}_{i=0, \ldots, K}$ is a monotonic non-increasing sequence.
\end{IEEEproof}

\section{Proof of Theorem \ref{Theorem_maxThroughput}} \label{Proof_Theorem_maxThroughput}
\begin{IEEEproof}
It is shown in (\ref{eq15}) that the network throughput can be obtained as an explicit function of $p_A$. The following lemma first presents $\hat{\lambda}^{p}_{\max}={\max}_{p_A\in(0,1]}\hat{\lambda}_{out}$ and the corresponding optimal steady-state point $p_A^{*}$.
\begin{lemma}\label{Lemma_maxThroughput}
For given SINR threshold $\mu\in(0,\infty)$ and mean received SNR $\rho\in(0,\infty)$, $\hat{\lambda}^{p}_{\max}$ is given by
\begin{equation}\label{maxThroughput}
\hat{\lambda}^{p}_{\max}=\tfrac{\mu+1}{\mu}\exp\left(-1-\tfrac{\mu}{\rho}\right),
\end{equation}
which is achieved at
\begin{equation}\label{maxThroughput_Pa}
p_A^*=\exp\left(-1-\tfrac{\mu}{\rho}\right).
\end{equation}
\end{lemma}
\begin{IEEEproof}
According to (\ref{eq15}), the second-order derivative of $\hat{\lambda}_{out}$ with respect to $p_A$ is given by $-\tfrac{\mu+1}{\mu p_A} <0$, for $p_A\in(0,\infty)$. Therefore, we can conclude that $\hat{\lambda}_{out}$ is a strictly concave function of $p_A\in (0,\infty)$ with one global maximum at $p_A^*$, where $p_A^*$ is the root of $\tfrac{d\hat{\lambda}_{out}}{dp_A}=0$, i.e., $(\mu+1)\cdot\left(\tfrac{-\ln p_A-1}{\mu}-\tfrac{1}{\rho} \right)=0$,
which is given by \eqref{maxThroughput_Pa}. Eq. \eqref{maxThroughput} can be obtained by substituting \eqref{maxThroughput_Pa} into \eqref{eq15}.
\end{IEEEproof}

We can see from Lemma \ref{Lemma_maxThroughput} and \eqref{Probability-of-success_4} that to achieve $\hat{\lambda}^{p}_{\max}$, the backoff parameters $\{q_i\}_{i=0, \ldots, K}$ should be carefully selected such that $p_A=p_A^*$. For given backoff function $\mathcal{Q}_i$, the optimal initial transmission probability for achieving $\hat{\lambda}^{p}_{\max}$ can be easily obtained by combining \eqref{Probability-of-success_4} and \eqref{maxThroughput_Pa} as \eqref{qm}.

Note that the initial transmission probability $q_0$ should not exceed $1$. Lemma \ref{Lemma_achievability} shows that $\hat{\lambda}^{p}_{\max}$ is achievable for $q_i\in(0,1]$, $i=0,\ldots,K$, if and only if the SINR threshold $\mu \geq \tfrac{1}{n-1}$.
\begin{lemma}\label{Lemma_achievability}
$\hat{\lambda}^{p}_{\max}$ is achievable if and only if $\mu \geq \tfrac{1}{n-1}$.
\end{lemma}
\begin{IEEEproof}
Define $\tilde {\mathcal{Q}}_i=1/{\mathcal{Q}}_i$, for $0\leq i\leq K-1$, and $\tilde {\mathcal{Q}}_i=1/\mathcal{Q}_K$ for $i\geq K$. Let $Y$ denote a geometric random variable with parameter $\exp\left(-1-\tfrac{\mu}{\rho}\right)$. Eq. \eqref{qm} can be then written as $\hat{q}_0=\tfrac{\mu+1}{n\mu}\cdot E_Y[\tilde {\mathcal{Q}}_Y]$.
As $\mathcal{Q}_i\leq1$ for $i=0, 1, \ldots, K$, we have $E_Y[\tilde {\mathcal{Q}}_Y]\geq1$.

\noindent 1) \textit{if}: if $\mu \geq \tfrac{1}{n-1}$, with $\mathcal{Q}_i=1$ for $i=0,1,\ldots,K$, we have $ E_Y[\tilde {\mathcal{Q}}_Y]=1$ and $\hat{q}_0=\tfrac{\mu+1}{n\mu}\leq 1$. In this case, $\hat{\lambda}^{p}_{\max}$ can be achieved by setting $q_0=\hat{q}_0$.

\noindent 2) \textit{only if}: if $\mu < \tfrac{1}{n-1}$, we have $\hat{q}_0\geq\tfrac{\mu+1}{n\mu} >1$, which indicates that $\hat{\lambda}^{p}_{\max}$ is not achievable.
\end{IEEEproof}

For $\mu< \tfrac{1}{n-1}$,  $\hat{\lambda}^{p}_{\max}$ is not achievable for $q_i\in(0,1]$, $i=0,\ldots,K$. The following lemma shows that in this case, the maximum network throughput $\hat{\lambda}_{\max}$ is always smaller than $\hat{\lambda}^{p}_{\max}$, which is achieved by setting $q_i=1$, $i=0,\ldots,K$.
\begin{lemma}\label{Lemma_maxThroughput_l}
For given SINR threshold $\mu< \tfrac{1}{n-1}$, the maximum network throughput $\hat{\lambda}_{\max}$ is given by
\begin{equation}\label{maxThroughput_l}
\hat{\lambda}_{\max}^{\mu<\tfrac{1}{n-1}}=n\exp\left(-\tfrac{n\mu}{\mu+1}-\tfrac{\mu}{\rho}\right),
\end{equation}
which is achieved at $q_i^*=1$, $i=0,\ldots,K$.
\end{lemma}
\begin{IEEEproof}
According to \eqref{Probability-of-success_4}, the initial transmission probability $q_0$ can be written as
\begin{equation}\label{Lemma_maxThroughput_l_eq1}
q_0=\tfrac{\mu+1}{n\mu}\left(-\ln p_A-\tfrac{\mu}{\rho}\right)\cdot z(p_A).
\end{equation}
where $z\left(p_A\right)=\sum_{i=0}^{K-1}\tfrac{p_A\left(1-p_A\right)^i}{\mathcal{Q}_i}+\tfrac{\left(1-p_A\right)^K}{\mathcal{Q}_K}$. Similar to $g(p)$ in Appendix \ref{Proof_Theorem_pA}, it can be proved that $z\left(p_A\right)$ is a monotonic non-increasing function of $p_A\in(0,1]$. Note that ${-}\ln p_A$ is also a monotonic non-increasing function of $p_A\in(0,1]$. Therefore, we can conclude from \eqref{Lemma_maxThroughput_l_eq1} that $p_A$ is a monotonic non-increasing function of $q_0$. With $0<q_0\leq1$, we can obtain from \eqref{Probability-of-success_4} that
\begin{equation}\label{Lemma_maxThroughput_l_eq2}
p_A\geq\exp\left(-\tfrac{\mu}{\rho}-\tfrac{n\mu}{\mu+1}\cdot \tfrac{1}{\resizebox{5mm}{!}{$\sum$}_{i=0}^{K-1}\tfrac{p_A\left(1-p_A\right)^i}{\mathcal{Q}_i}+\tfrac{\left(1-p_A\right)^K}{\mathcal{Q}_K}}\right),
\end{equation}
where ``$=$'' holds when $q_0=1$. Note that the backoff function $\mathcal{Q}_i\leq1$, $i=0,\ldots,K$. We can further obtain from \eqref{Lemma_maxThroughput_l_eq2} that
\begin{equation}\label{Lemma_maxThroughput_l_eq3}
p_A\geq\exp\left(-\tfrac{\mu}{\rho}-\tfrac{n\mu}{\mu+1}\right),
\end{equation}
where ``$=$'' holds when $q_0=1$ and $\mathcal{Q}_i=1$, $i=0,\ldots,K$.

When $\mu<\tfrac{1}{n-1}$, we can see from \eqref{Lemma_maxThroughput_l_eq3} that $p_A>\exp\left(-\tfrac{\mu}{\rho}-1\right)=p_A^*$. According to the proof of Lemma \ref{Lemma_maxThroughput}, the network throughput $\hat{\lambda}_{out}$ is a monotonic decreasing function of $p_A$ when $p_A>p_A^*$. Therefore, in this case, $\hat{\lambda}_{out}$ is maximized when $p_A$ is minimized, i.e., $p_A=\exp\left(-\tfrac{\mu}{\rho}-\tfrac{n\mu}{\mu+1}\right)$ according to \eqref{Lemma_maxThroughput_l_eq3}, which is achieved at $q_i^*=1$. Eq. \eqref{maxThroughput_l} can be then obtained by substituting $p_A=\exp\left(-\tfrac{\mu}{\rho}-\tfrac{n\mu}{\mu+1}\right)$ into \eqref{eq15}.
\end{IEEEproof}

Finally, Eqs. (\ref{maxThroughput_all}) and (\ref{maxThroughput_q}) can be obtained by combining Lemma \ref{Lemma_maxThroughput}, Lemma \ref{Lemma_achievability}  and Lemma \ref{Lemma_maxThroughput_l}.
\end{IEEEproof}

\section{Proof of Theorem \ref{Theorem_maximumRate}} \label{Proof_Theorem_maximumRate}
\begin{IEEEproof}
According to (\ref{maxRate_f}), we can rewrite the maximum sum rate as $C=\max\left(C_1,C_2\right)$, where
\begin{equation}\label{Proof_Theorem_maximumRate_eq1}
C_1=\max_{\mu\geq \tfrac{1}{n-1}} \tfrac{\mu+1}{\mu}\exp\left (-1-\tfrac{\mu}{\rho}\right)\cdot \log_2(1+\mu),
\end{equation}
and
\begin{equation}\label{Proof_Theorem_maximumRate_eq2}
C_2=\max_{0<\mu\leq \tfrac{1}{n-1}} n\exp\left (-\tfrac{n\mu}{\mu+1}-\tfrac{\mu}{\rho}\right)\cdot \log_2(1+\mu).
\end{equation}
Let us first focus on $C_1$.

\noindent 1) Denote the objective function of \eqref{Proof_Theorem_maximumRate_eq1} as $f_1(\mu)$ and let us first prove the following lemma.
\begin{lemma}\label{Lem_objectiveFunction-of-sumRate}
$f_1(\mu)$ is a monotonic decreasing function of $\mu\in\left[\tfrac{1}{n-1},\infty\right)$ if $\rho<\rho_0$. Otherwise, it has one global maximum at $\mu_h^*$, where $\mu_h^*$ is the root of \eqref{optimal-SINR-threshold_root_h}.
\end{lemma}
\begin{IEEEproof}
$f_1(\mu)$ is a continuously differentiable function of $\mu\in\left[\tfrac{1}{n-1},\infty\right)$. The first-order derivative of $f_1(\mu)$ can be written as
\begin{equation}\label{Lemma1_eq1}
f_1'(\mu)=\exp\left(-1-\tfrac{\mu}{\rho}\right)\log_2 e\cdot G_1(\mu),
\end{equation}
where
\begin{equation}\label{G_1}
G_1(\mu)=\tfrac{1}{\mu}-\tfrac{1}{\mu^2}\ln (1+\mu)-\tfrac{1}{\rho}\cdot\tfrac{1+\mu}{\mu}\ln(1+\mu).
\end{equation}
It can be easily obtained from \eqref{G_1} that
\begin{equation}\label{G_1-Mu-to-0}
\lim_{\mu \to \tfrac{1}{n-1}} G_1(\mu)=(n-1)-(n-1)^2\ln \tfrac{n}{n-1}-\tfrac{n}{\rho}\ln \tfrac{n}{n-1},
\end{equation}
and
\begin{equation}\label{G_1-Mu-to-Inf}
\lim_{\mu \to \infty} G_1(\mu)=-\infty.
\end{equation}
Moreover, the first-order derivative of $G_1(\mu)$ can be obtained from \eqref{G_1} as
\begin{equation} \label{G_1-First_order}
G_1'(\mu)=-\tfrac{1}{\mu^2}\left(\tfrac{2+\mu}{1+\mu}-\tfrac{2}{\mu}\ln(1+\mu)\right)-\tfrac{1}{\rho}\left(\tfrac{1}{\mu}-\tfrac{\ln(1+\mu)}{\mu^2}\right)<0,
\end{equation}
for $\mu\in\left[\tfrac{1}{n-1},\infty\right)$, which indicates that $G_1(\mu)$ is a monotonic decreasing function of $\mu\in\left[\tfrac{1}{n-1},\infty\right)$.

\noindent i) If $\rho\geq\rho_0$, we can obtain from (\ref{G_1-Mu-to-0}-\ref{G_1-Mu-to-Inf}) that $\lim_{\mu\rightarrow\tfrac{1}{n-1}}G_1(\mu)\geq0$ and $\lim_{\mu \to \infty} G_1(\mu)<0$. As $G_1(\mu)$ is a monotonic decreasing function of $\mu\in\left[\tfrac{1}{n-1},\infty\right)$, there must exist $\mu_h^*\in\left[\tfrac{1}{n-1},\infty\right)$, such that $G_1(\mu)>0$ for $\mu\in\left[\tfrac{1}{n-1},\mu_h^*\right)$ and $G_1(\mu)<0$ for $\mu\in\left(\mu_h^*,\infty\right)$, where $\mu_h^*$ is the root of $G_1(\mu)=0$, which is given in \eqref{optimal-SINR-threshold_root_h}. We can then obtain from \eqref{Lemma1_eq1} that $f_1'(\mu)>0$ for $\mu\in\left[\tfrac{1}{n-1},\mu_h^*\right)$ and $f_1'(\mu)<0$ for $\mu\in\left(\mu_h^*,\infty\right)$, which indicates that $f_1(\mu)$ has one global maximum at $\mu_h^*$.

\noindent ii) If $\rho<\rho_0$, we can obtain from \eqref{G_1-Mu-to-0} that $\lim_{\mu\rightarrow \tfrac{1}{n-1}}G_1(\mu)<0$. As $G_1(\mu)$ is a monotonic decreasing function of $\mu\in\left[\tfrac{1}{n-1},\infty\right)$, we have $G_1(\mu)<0$ for $\mu\in\left[\tfrac{1}{n-1},\infty\right)$. According to \eqref{Lemma1_eq1}, we can conclude that in this case $f_1(\mu)$ is a monotonic decreasing function as $f_1'(\mu)<0$ for $\mu\in\left[\tfrac{1}{n-1},\infty\right)$.
\end{IEEEproof}
According to Lemma \ref{Lem_objectiveFunction-of-sumRate}, we can conclude that the optimal SINR threshold for $C_1$ is
\begin{equation}\label{Proof_Theorem_maximumRate_opt1}
\mu_1^*=\begin{cases}
\mu_h^* & \text{if\;}\rho\geq\rho_0 \\
\tfrac{1}{n-1} & \text{otherwise}.
\end{cases}
\end{equation}
2) For $C_2$, denote the objective function of \eqref{Proof_Theorem_maximumRate_eq2} as $f_2(\mu)$ and let us first prove the following lemma.
\begin{lemma}\label{Lem_objectiveFunction-of-sumRate2}
$f_2(\mu)$ is a monotonic non-decreasing function of $\mu\in\left(0,\tfrac{1}{n-1}\right]$ if $\rho\geq\rho_0$. Otherwise, it has one global maximum at $\mu_l^*$, where $\mu_l^*$ is the root of \eqref{optimal-SINR-threshold_root_l}.
\end{lemma}
\begin{IEEEproof}
$f_2(\mu)$ is a continuously differentiable function of $\mu\in\left(0,\tfrac{1}{n-1}\right]$. The first-order derivative of $f_2(\mu)$ can be written as
\begin{equation}\label{Lemma1_eq2}
f_2'(\mu)=\tfrac{n}{(1+\mu)^2}\exp\left(-\tfrac{n\mu}{\mu+1}-\tfrac{\mu}{\rho}\right)\log_2 e \cdot G_2(\mu),
\end{equation}
where
\begin{equation}\label{Proof_Theorem_maximumRate_G3}
G_2(\mu)=(1+\mu)-\left(\tfrac{(1+\mu)^2}{\rho}+n\right)\ln(1+\mu).
\end{equation}
It can be easily obtained from \eqref{Proof_Theorem_maximumRate_G3} that
\begin{equation}\label{Proof_Theorem_maximumRate_G3_1}
\lim_{\mu\rightarrow 0}G_2(\mu)=1,
\end{equation}
and
\begin{equation}\label{Proof_Theorem_maximumRate_G3_2}
\lim_{\mu\rightarrow \tfrac{1}{n-1}}G_2(\mu)=\tfrac{n}{n-1}-n\ln\tfrac{n}{n-1}-\tfrac{1}{\rho}\cdot\left(\tfrac{n}{n-1}\right)^2 \ln\tfrac{n}{n-1}.
\end{equation}
Moreover, the first-order derivative of $G_2(\mu)$ can be obtained from \eqref{Proof_Theorem_maximumRate_G3} as
\begin{equation}\label{Proof_Theorem_maximumRate_G3_3}
G_2'(\mu)=1-\tfrac{n}{1+\mu}-\tfrac{1+\mu}{\rho}\left(1+2\ln(1+\mu)\right)<0,
\end{equation}
for $\mu\in\left(0,\tfrac{1}{n-1}\right]$, which indicates that $G_2(\mu)$ is a monotonic decreasing function of $\mu\in\left(0,\tfrac{1}{n-1}\right]$.

\noindent i) If $\rho\geq\rho_0$, we can obtain from \eqref{Proof_Theorem_maximumRate_G3_2} that $\lim_{\mu\rightarrow\tfrac{1}{n-1}}G_2(\mu)\geq0$. As $G_2(\mu)$ is a monotonic decreasing function of $\mu\in(0,\tfrac{1}{n-1}]$, we have $G_2(\mu)\geq0$ for $\mu\in\left(0,\tfrac{1}{n-1}\right]$. According to \eqref{Lemma1_eq2}, we can conclude that in this case $f_2(\mu)$ is a monotonic non-decreasing function as $f_2'(\mu)\geq0$ for $\mu\in\left(0,\tfrac{1}{n-1}\right]$.

\noindent ii) If $\rho<\rho_0$, we can obtain from (\ref{Proof_Theorem_maximumRate_G3_1}-\ref{Proof_Theorem_maximumRate_G3_2}) that $\lim_{\mu\rightarrow 0}G_2(\mu)>0$ and $\lim_{\mu\rightarrow \tfrac{1}{n-1}}G_2(\mu)<0$. As $G_2(\mu)$ is a monotonic decreasing function of $\mu\in\left(0,\tfrac{1}{n-1}\right]$, there must exist $\mu_l^*\in\left(0,\tfrac{1}{n-1}\right]$, such that $G_2(\mu)>0$ for $\mu\in(0,\mu_l^*)$ and $G_2(\mu)<0$ for $\mu\in\left(\mu_l^*,\tfrac{1}{n-1}\right]$, where $\mu_l^*$ is the root of $G_2(\mu)=0$, which is given in \eqref{optimal-SINR-threshold_root_l}. We can then obtain from \eqref{Lemma1_eq2} that $f_2'(\mu)>0$ for $\mu\in(0,\mu_l^*)$ and $f_2'(\mu)<0$ for $\mu\in\left(\mu_l^*,\tfrac{1}{n-1}\right]$, which indicates that $f_2(\mu)$ has one global maximum at $\mu_l^*$.
\end{IEEEproof}
According to Lemma \ref{Lem_objectiveFunction-of-sumRate2}, we can conclude that the optimal SINR threshold for $C_2$ is
\begin{equation}\label{Proof_Theorem_maximumRate_opt2}
\mu_2^*=\begin{cases}
\tfrac{1}{n-1} & \text{if\;}\rho\geq\rho_0 \\
\mu_l^*& \text{otherwise}.
\end{cases}
\end{equation}

\noindent 3) By combining \eqref{Proof_Theorem_maximumRate_opt1} and \eqref{Proof_Theorem_maximumRate_opt2}, we can see that if $\rho\geq\rho_0$, $C_1=f_1(\mu_h^*)$ and $C_2=f_2\left(\tfrac{1}{n-1}\right)$. As $f_2\left(\tfrac{1}{n-1}\right)=f_1\left(\tfrac{1}{n-1}\right)$ and $f_1\left(\tfrac{1}{n-1}\right)\leq C_1$, we have $C_1\geq C_2$. Therefore, we can conclude that in this case the maximum sum rate $C=C_1$ and the optimal SINR threshold $\mu^*=\mu_h^*$.

On the other hand, if $\rho<\rho_0$, $C_1=f_1\left(\tfrac{1}{n-1}\right)$ and $C_2=f_2(\mu_l^*)$. As $f_1\left(\tfrac{1}{n-1}\right)=f_2\left(\tfrac{1}{n-1}\right)$ and $f_2\left(\tfrac{1}{n-1}\right)\leq f_2(\mu_l^*)$, we have $C_2\geq C_1$. Therefore, we can conclude that in this case the maximum sum rate $C=C_2$ and the optimal SINR threshold $\mu^*=\mu_l^*$.
\end{IEEEproof}

\section{Proof of Corollary \ref{Corollary_highSNR}} \label{Proof_Corollary_highSNR}
\begin{IEEEproof}
When $\rho\geq\rho_0$, the optimal SINR threshold $\mu^*=\mu_h^*$ according to \eqref{optimal-SINR-threshold}. We can easily obtain from \eqref{optimal-SINR-threshold_root_h} that $\lim_{\rho\to\infty}\mu_h^{*}=\infty$, and
\begin{equation}\label{COROLLARY_lim_muo}
\lim_{\rho\to\infty} \tfrac{\mu_h^{*}}{\rho}\ln \mu_h^{*}= \lim_{\rho\to\infty} \left(
\tfrac{1}{\mu_h^{*}}+\tfrac{1+\mu_h^{*}}{\rho}\right)\ln(1+\mu_h^{*})=1.
\end{equation}
According to (\ref{COROLLARY_lim_muo}), we further have
\begin{equation}\label{COROLLARY_lim_mu-rho}
\lim_{\rho\to\infty} \tfrac{\mu_h^{*}}{\rho} =\lim_{\rho\to\infty} \tfrac{1}{\ln \mu_h^{*}}=0.
\end{equation}
Moreover, by applying L'H\^{o}pital's rule on the left-hand side of (\ref{COROLLARY_lim_muo}), we can obtain that
\begin{equation}\label{COROLLARY_dmu}
\lim_{\rho\to\infty} \tfrac{d\mu_h^{*}}{d\rho}(1+\ln\mu_h^{*})=1.
\end{equation}
Finally, by combining \eqref{maxRate_2} with (\ref{COROLLARY_lim_muo}-\ref{COROLLARY_dmu}), we have $\lim_{\rho\to\infty} \tfrac{C}{\log_2 \rho}=e^{-1} \cdot \lim_{\rho\to\infty} \tfrac{\log_2 \mu_h^*}{\log_2 \rho}=e^{-1} \cdot \lim_{\rho\to\infty} \tfrac{\rho}{\mu_h^*(1+\ln \mu_h^*)}=e^{-1}$.
\end{IEEEproof}

\section{Proof of Corollary \ref{Corollary_lowSNR}} \label{Proof_Corollary_lowSNR}
\begin{IEEEproof}
When $\rho<\rho_0$, the optimal SINR threshold $\mu^*=\mu_l^*$ according to \eqref{optimal-SINR-threshold}. We can easily obtain from \eqref{optimal-SINR-threshold_root_l} that $\lim_{n\rightarrow \infty} \mu_l^{*}=0$, and
\begin{equation}\label{Proof_Corollary_lowSNR_eq2}
\lim_{n\rightarrow \infty}n\log_2(1+\mu_l^{*})=\lim_{n\rightarrow \infty} \tfrac{\log_2 e}{\tfrac{1}{\mu_l^*+1}+\tfrac{\mu_l^*+1}{n\rho}}=\log_2 e,
\end{equation}
\begin{equation}\label{Proof_Corollary_lowSNR_eq3}
\lim_{n\rightarrow \infty}\tfrac{n\mu_l^*}{\mu_l^{*}+1}=\lim_{n\rightarrow \infty} \tfrac{\mu_l^*}{\ln(1+\mu_l^*)}-\tfrac{\mu_l^*(\mu_l^*+1)}{\rho}=1.
\end{equation}
Finally, by combining \eqref{maxRate_2} with (\ref{Proof_Corollary_lowSNR_eq2}-\ref{Proof_Corollary_lowSNR_eq3}), we have $\lim_{n\rightarrow\infty}C_{\rho<\rho_0}=\lim_{n\rightarrow\infty} n\exp\left(-\tfrac{n\mu_l^{*}}{\mu_l^{*}+1}-\tfrac{\mu_l^{*}}{\rho}\right)\cdot\log_2(1+\mu_l^{*})=e^{-1}\log_2 e$.
\end{IEEEproof}

\begin{IEEEbiography}[{\includegraphics[width=1in,height=1.25in,clip,keepaspectratio]{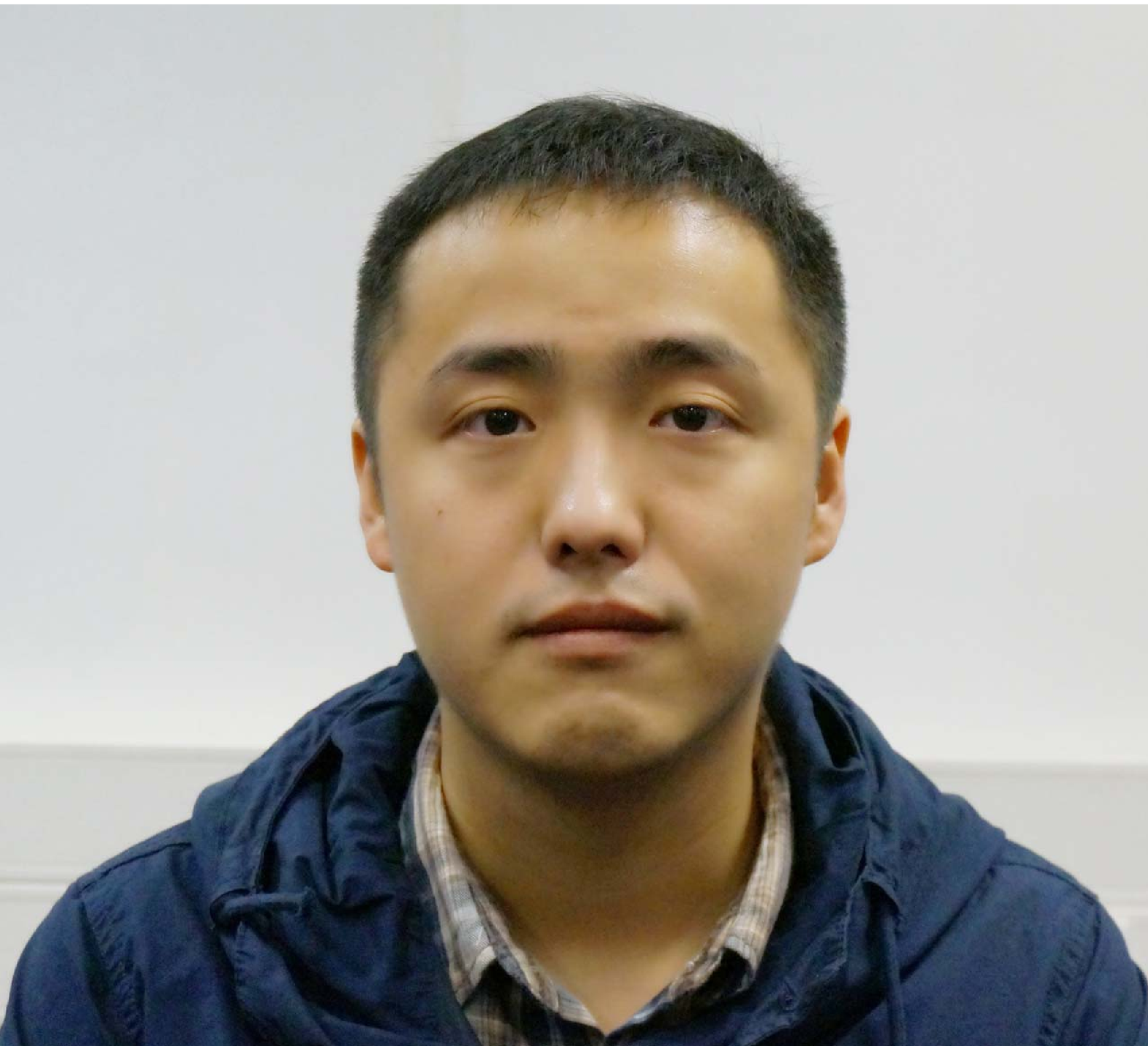}}]{Yitong Li}
received the B.Eng. degree in Electronic Engineering in 2011 from City University of Hong Kong, Kowloon, Hong Kong, where he is currently working toward the Ph.D. degree with the Department of Electronic Engineering. His research interests include performance evaluation and optimization of wireless random access networks.
\end{IEEEbiography}

\begin{IEEEbiography}[{\includegraphics[width=1in,height=1.25in,clip,keepaspectratio]{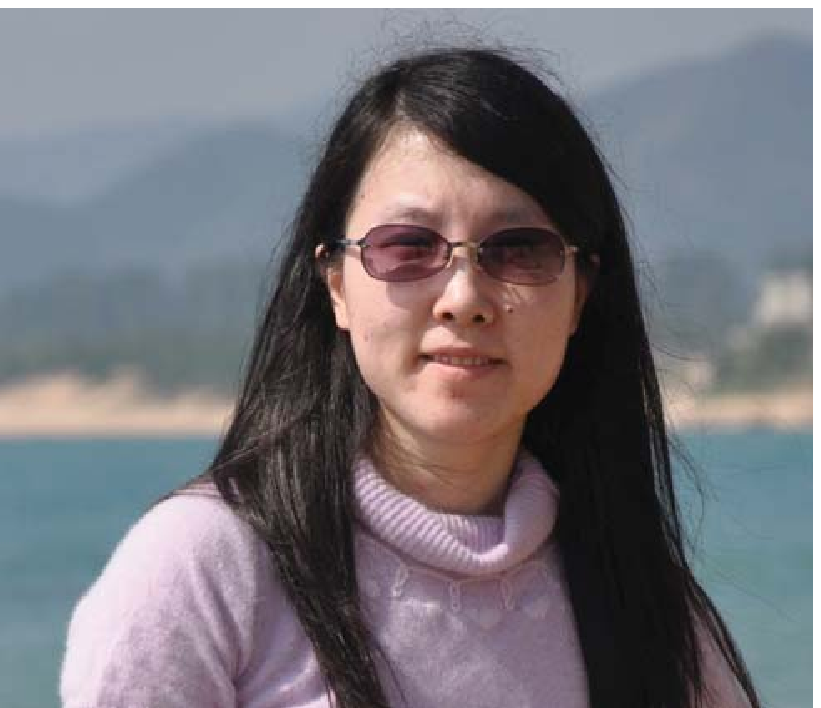}}]{Lin Dai}
(S'00-M'03-SM'13) received the B.S. degree from Huazhong University of Science and Technology, Wuhan, China, in 1998, and the M.S. and Ph.D. degrees from Tsinghua University, Beijing, China, in 2003, all in electronic engineering.

She was a postdoctoral fellow at The Hong Kong University of Science and Technology and University of Delaware. Since 2007, she has been with City University of Hong Kong, where she is an associate professor. She has broad interest in communications and networking theory, with special interest in wireless communications. She was a co-recipient of the Best Paper Award at the IEEE Wireless Communications and Networking Conference (WCNC) 2007 and the IEEE Marconi Prize Paper Award in 2009.
\end{IEEEbiography}
\vfill

\end{document}